\documentclass[a4paper,superscriptaddress,twocolumn,pra,showkeys,preprintnumbers,floatfix,showpacs]{revtex4-1}

\usepackage{epsfig}
\usepackage{amsmath}
\usepackage{amssymb}
\usepackage{grffile}
\usepackage{color}
\usepackage[dvipsnames]{xcolor}
\usepackage{ulem}

\newcommand{\beq}{\begin{eqnarray*}}
\newcommand{\eeq}{\end{eqnarray*}}

\newcommand{\be}{\begin{eqnarray}}
\newcommand{\ee}{\end{eqnarray}}

\usepackage[colorlinks,bookmarks=false,citecolor=blue,linkcolor=purple,urlcolor=blue]{hyperref}

\begin{document}
\title{Suppression of the horizon effect in pairing correlation functions of $t$-$J$ chains after a quantum quench}

\author{A. K\"uhn}
\affiliation{Institut f\"ur Theoretische Physik, Georg-August-Universit\"at G\"ottingen, 37077 G\"ottingen, Germany}

\author{L. Cevolani}
\affiliation{Institut f\"ur Theoretische Physik, Georg-August-Universit\"at G\"ottingen, 37077 G\"ottingen, Germany}


\author{S.R. Manmana}
\affiliation{Institut f\"ur Theoretische Physik, Georg-August-Universit\"at G\"ottingen, 37077 G\"ottingen, Germany}

\date{\today}

\begin{abstract}
We investigate the time evolution of density, spin, and pairing correlation functions in one-dimensional $t$-$J$ models following a quantum quench using the time-dependent density matrix renormalization group (tDMRG). 
While density and spin correlation functions show the typical light-cone behavior over a wide range of parameters, in pairing correlation functions it is strongly suppressed.
This is supported by time-dependent BCS theory, where the light-cone in the pairing correlation functions is found to be at least two orders of magnitude weaker than in the density correlator. 
These findings indicate that in global quantum quenches not all observables are affected equally by the excitations induced by the quench.
\end{abstract}

\maketitle

\section{Introduction}
The spread of correlations and information in a quantum lattice system is constrained by the existence of a maximal speed. 
This was proven for the commutator of observables for short-ranged systems by Lieb and Robinson~\cite{LiebRobinson72}, and later also for correlation functions~\cite{bravyi2006lieb,Kastner_NJP}.
This corresponds to the findings in global quantum quenches, where light-cone behavior in correlation functions has been reported in theoretical investigations (see, e.g.,~\cite{calabrese06,calabrese07,laeuchli08,manmana09,Carleo2014}),  
as well as in experiments with ultracold gases on optical lattices \cite{Cheneau:2012bh,lightcone_Langen2013}. 
As pointed out by Calabrese and Cardy~\cite{calabrese06,calabrese07}, a way to picture such an horizon effect is to see the quench as a local source for quasi-particles, which then move ballistically through the system, i.e., at a constant speed.
They leave the typical linear signature in the time evolution of correlation functions, which travels with twice the velocity of the quasi-particles, since it is caused by two of them moving in opposite directions. 
An interesting question is if this behavior is typical for all quantum systems, and one direction considered in recent research is to treat systems with long-range interactions~\cite{Hastings2006,PRA_Correlations,PRL_Eisert,PRL_ZheXuan, Hauke2013, Cevolani2015, Cevolani2016, Cevolani2017, Frerot2018, Vodola2014, Foss-Feig2015, Richerme2014, Schachenmayer2015, Buyskikh2016, Jurcevic2014, Regemortel2016}.

Here, we ask the question whether the light-cone signal is expected to behave generically in all propagators.
In Ref.~\onlinecite{PhysRevLett.107.115301} it is reported that in the center of mass motion (
) the frequency of oscillations doubles when entering a Luther-Emery like phase \cite{giamarchi} with dominant singlet pairing correlation functions in a one-dimensional $t$-$J$ model.
For the sake of simplicity, in the following we will refer to this phase as 'superconducting' (SC), although according to the Mermin-Wagner-Hohenberg theorem \cite{merminwagner_orig,merminwagner_erratum,hohenberg_orig} no true formation of pairs is possible.
This observation on the COM indicates that SC quasi-long-rang-order can affect the time evolution of observables on a qualitative level. 
Here, we want to investigate if such SC phases can have an effect on the time evolution of further observables, like correlation functions.
The scope of this paper is to address this question by considering quantum quenches in a variant of the $t$-$J$ model with only transverse spin interactions in one spatial dimension, which has such a SC phase of significant size \cite{tJperp_PRA}.
This model realizes a quantum simulator for magnetism and superconductivity using ultracold polar molecules  \cite{reviewmolecules,Lemeshko_Review,PhysRevLett.101.133004,Silke_science,Silke_science2,Silke_Nature,
PhysRevLett.104.030402,
Amodsen,PhysRevLett.84.246,PhysRevLett.94.203001,PhysRevLett.101.133005,deMiranda:2011gd,reviewmolecules,
PhysRevLett.112.070404,focus_ultracoldmolecules,PhysRevLett.114.205302,
PhysRevLett.116.225306,review_molecules2,MolonyPRL,TakekoshiPRL,GoulvenPRL2016,Yan:2013fn} on optical lattices~\cite{Bloch:2005p988,Bloch:2008p943}.

In this paper, we focus on one-dimensional systems, for which one can efficiently compute the time evolution following a quantum quench by using the time dependent density matrix renormalization group (tDMRG) \cite{white1992,white1993,dmrgbook,daley04,white04,feiguin05,manmana:269,noack:93,Schollwock:2011p2122}.
The main finding is that the light-cone is strongly suppressed or absent in pairing correlation functions.
To further elucidate this observation, we compute the time evolution of the correlation functions also using a BCS approach. 
The main results of this mean-field approach agree well with the results of the tDMRG.
In particular, the BCS theory shows that the amplitude of the light-cone in the pairing correlation functions is strongly suppressed compared to the one in the density-density correlators.
Similar behavior is also found in the BCS-treatment of two-dimensional systems. 
As this approach is independent of the choice of the microscopic lattice model, this gives strong indications for the light-cone in pairing correlation functions to be generally suppressed in the time evolution following a global quantum quench.

The paper is organized as follows: In Sec.~\ref{sec:model} we introduce the variant of the $t$-$J$ model and the observables treated in this paper.
In Sec.~\ref{sec:results} we present our tDMRG results for the local densities and for the correlation functions.
In Sec.~\ref{sec:BCS} we discuss our BCS theory and its main results for the density-density and pairing correlation functions.
Sec.~\ref{sec:conclusions} provides a summary.
Appendix~\ref{sec:appendix_BCS} contains details to the BCS calculations, and Appendix~\ref{app:2dmean} exemplifies the mean-field results for the simplest two-dimensional case.  

\section{Model and Observables}
\label{sec:model}

\subsection{$t$-$J$ Model with Transverse Spin Exchange Interactions}

In Refs.~\onlinecite{PhysRevLett.107.115301,PhysRevA.84.033619} the Hamiltonian 
\begin{equation}
\begin{split}
&\mathcal H^{tJVW} =  - t_{\rm hop} \sum_{i,\sigma}  \left[c^\dagger_{i,\sigma} c^{\phantom{\dagger}}_{i+1,\sigma} + \textrm{h.c.}\right] \\
&+ \sum_{j>i} \frac{1}{|i-j|^3} \left[  \frac{J_\perp}{2} \left( S^+_i S^-_j  + S^-_i S^+_j \right) + J_z S^z_{i} S^z_{j} \right. \\
& \left. \phantom{\frac{J_\perp}{2}} + V n^{\phantom{\dagger}}_i n^{\phantom{\dagger}}_j + W n^{\phantom{z}}_i S^z_{j}  \right]  
\end{split}
\label{eq:tjvw}
\end{equation}
is derived as quantum simulator for quantum magnetism and superconductivity in systems of ultracold polar molecules~\cite{reviewmolecules,Lemeshko_Review,PhysRevLett.101.133004,Silke_science,Silke_science2,Silke_Nature,
PhysRevLett.104.030402,
Amodsen,PhysRevLett.84.246,PhysRevLett.94.203001,PhysRevLett.101.133005,deMiranda:2011gd,reviewmolecules,
PhysRevLett.112.070404,focus_ultracoldmolecules,PhysRevLett.114.205302,
PhysRevLett.116.225306,review_molecules2,MolonyPRL,TakekoshiPRL,GoulvenPRL2016} on optical lattices~\cite{Bloch:2005p988,Bloch:2008p943}. 
The operators $c^{(\dagger)}_{i,\sigma}$ are fermionic annihilation (creation) operators for a particle with spin $\sigma$ on lattice site $i$, the Hilbert space is the usual fermionic Hilbert space projected onto the space with no doublons (as in the usual $t$-$J$ model), $S^+_i = c^\dagger_{i,\uparrow}c^{\phantom{\dagger}}_{i,\downarrow}$ and $S^-_i = c^\dagger_{i,\downarrow}c^{\phantom{\dagger}}_{i,\uparrow}$ are the spin raising and lowering operators, $S^z_i = (c^\dagger_{i,\uparrow} c^{\phantom{\dagger}}_{i,\uparrow} - c^\dagger_{i,\downarrow} c^{\phantom{\dagger}}_{i,\downarrow} )/2$ is the $z$ component of the spin operator, and $n^{\phantom{\dagger}}_i = \sum_{\sigma} c^\dagger_{i,\sigma} c^{\phantom{\dagger}}_{i,\sigma}$ is the total density on site $i$. 
In Ref.~\onlinecite{Yan:2013fn} it is found that spin-exchange interactions are indeed realized in such experiments, paving the way for further developments.

Model~\eqref{eq:tjvw} is a generalization of the standard $t$-$J$ model \cite{tJoriginal1,tJoriginal2,auerbach,dagotto,tJ1977}, which in one dimension (1D) reads 
\begin{equation}
\begin{split}
\mathcal H ^{tJ} = - t_{\rm hop} \sum_{i,\sigma}& \left[ c_{i,\sigma}^{\dagger} c_{i+1,\sigma}^{\phantom{\dagger}} + \textrm{h.c.} \right] \\
& + J \sum_i \left[ \vec{S}^{\phantom{\dagger}}_i \cdot \vec{S}^{\phantom{\dagger}}_{i+1} - \frac{1}{4} n^{\phantom{\dagger}}_i n^{\phantom{\dagger}}_{i+1} \right], 
\end{split}
\label{eq:standard_tJ}
\end{equation}
and which usually is obtained via second-order degenerate perturbation theory from the Hubbard model~\cite{auerbach}. 
In perturbation theory, one finds $J = 4t_{\rm hop}^2/U$, with $U$ the strength of the Hubbard interaction, and it is not possible to tune the parameters $t,\, J_\perp, \, J_z, \, V$ and $W$ independently from each other. 
In contrast, in the polar molecules setup \cite{PhysRevLett.107.115301,PhysRevA.84.033619} the values of the parameters are fully tunable.  
Note that model~\eqref{eq:standard_tJ} is obtained from Eq.~\eqref{eq:tjvw} by considering only nearest neighbor interactions and setting $J_z = J_\perp \equiv J$, $V = -J/4$ and $W = 0$.
Here, we treat model~\eqref{eq:tjvw} with $J_z = V = W = 0$ and for simplicity we consider only nearest neighbor interactions, leading us to the $t$-$J_\perp$ chain
\begin{equation}
\begin{split}
\mathcal H^{tJ_\perp} = &- t_{\rm hop} \sum_{i,\sigma} \left[ c_{i,\sigma}^{\dagger} c_{i+1,\sigma}^{\phantom{\dagger}} + \textrm{h.c.} \right] \\
&+ \frac{J_\perp}{2} \sum_{i} \left[ S_i^+ S_{i+1}^- + S_i^- S_{i+1}^+ \right] \, . 
\end{split}
\label{eq:tJperp}
\end{equation}
The ground state phase diagram of model~\eqref{eq:tJperp} in one spatial dimension is discussed in Refs.~\onlinecite{PhysRevLett.107.115301,tJperp_PRA} and is found to be similar to the one of model~\eqref{eq:standard_tJ} discussed in Ref.~\onlinecite{Moreno}.
An important difference is that the singlet superconducting (SC) phase at low fillings is significantly enhanced.
As in this paper we want to study the effect of a SC phase on the horizon effect, this is favorable, since it allows us to stay farther away from phase transition points, so that their possible effect would have less influence on the dynamics.  

\subsection{Observables}

We treat the following observables in this paper. 
First, we consider the behavior of on-site quantities like the local density $n_i$ and the local magnetizations $S^z_i$ on lattice site $i$.
We consider systems with zero total magnetization $\langle \sum_i S^z_i \rangle = 0$.
According to the Mermin-Wagner-Hohenberg theorem \cite{merminwagner_orig,merminwagner_erratum,hohenberg_orig}, also the local magnetizations are zero, since the corresponding continuous symmetry cannot be broken spontaneously in one spatial dimension.
This is found to be true also in the course of the time evolution.
 
The connected density-density correlation function is given by
\begin{equation}
N_{ij} = \langle n_i n_j \rangle - \langle n_i \rangle \langle n_j \rangle \, .
\label{eq:densdens}
\end{equation}
For the spin correlation functions, one in principle has to treat the longitudinal component 
\begin{equation}
C^{\rm spin, long}_{ij} = \langle S_i^z S_j^z \rangle  - \langle S_i^z \rangle \langle S_j^z \rangle 
\label{eq:SzSz}
\end{equation}
and the transverse component
\begin{equation}
C^{\rm spin, trans}_{ij} = \langle S_i^+ S_j^- \rangle
\label{eq:Strans}
\end{equation}
independently, as the $t$-$J_\perp$ model lacks SU(2) invariance.
However, since the features in the time evolution are very similar, we will mainly discuss $C^{\rm spin, long}_{ij}$.

The superconducting properties are probed by the pairing correlation functions
\begin{equation}
P^{T,S}_{ij} = \langle \Delta^{\dagger}_{T,S}(i) \, \Delta^{\phantom \dagger}_{T,S}(j) \rangle.
\label{eq:pairing}
\end{equation}
In this paper we treat 
\begin{equation}
\Delta^{\dagger}_S(i) = \frac{1}{\sqrt{2}} \left(c_{i,\downarrow} ^\dagger c_{i+1,\uparrow} ^\dagger - c_{i,\uparrow} ^\dagger c_{i+1,\downarrow} ^\dagger \right)
\label{eq:delta_singlet}
\end{equation}
for singlet pairing and
\begin{equation}
\Delta^{\dagger}_T(i) = c_{i,\uparrow} ^\dagger c_{i+1,\uparrow} ^\dagger
\label{eq:delta_triplet}
\end{equation}
for triplet pairing. 

Throughout the paper we work in units, in which $\hbar \equiv 1 $ and we set $t_{\rm hop} = 1$.

\subsection{Details for the time-dependent DMRG}
\label{subsec:DMRG}

We obtain the ground state using the density matrix renormalization group (DMRG)~\cite{white1992,white1993,dmrgbook,noack:93,Schollwock:2005p2117,
Schollwock:2011p2122} performing 5 sweeps and keeping up to $m=1000$ states. 
We treat systems with $N=16$ particles on $L=80$ lattice sites, corresponding to a filling of 0.2, and zero total magnetization.
We apply the adaptive time-dependent DMRG method using a Trotter time evolution scheme \cite{vidal03,daley04,white04} for computing the dynamics of the system~\eqref{eq:tJperp}.
The time evolution is initiated by a quantum quench, in which we keep $t_{\rm hop}=1$ fixed and suddenly change the value of $J_\perp$.
We keep up to $m=1400$ states during the time evolution and find in the worst case a discarded weight $\sim 10^{-8}$ at the maximal time explored.  
In all cases, we apply open boundary conditions (OBC).  

\section{Local Observables and Correlation Functions}
\label{sec:results}

\subsection{Local density}
\begin{figure}[b]
\includegraphics[width=\columnwidth]{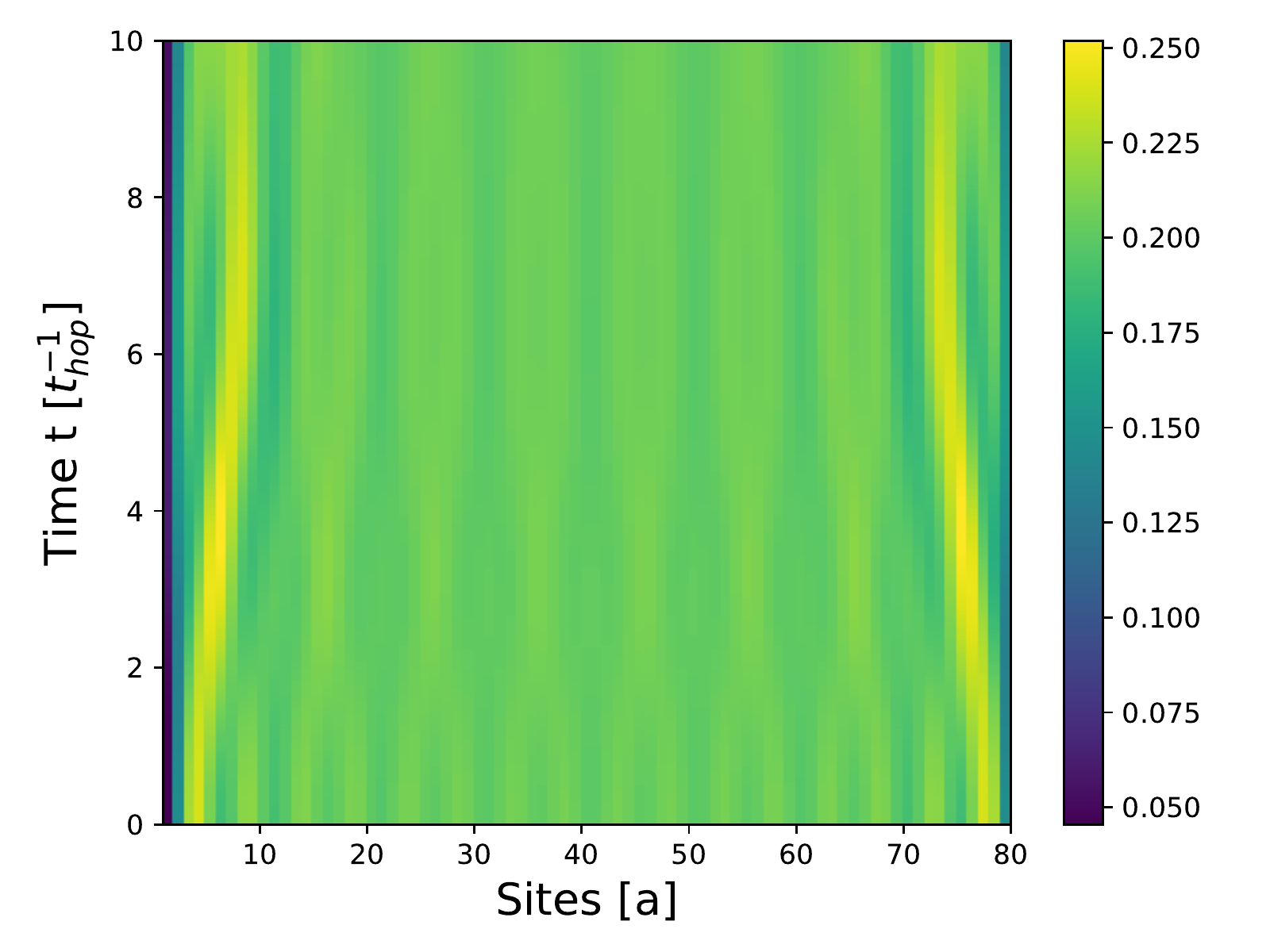}
\includegraphics[width=\columnwidth]{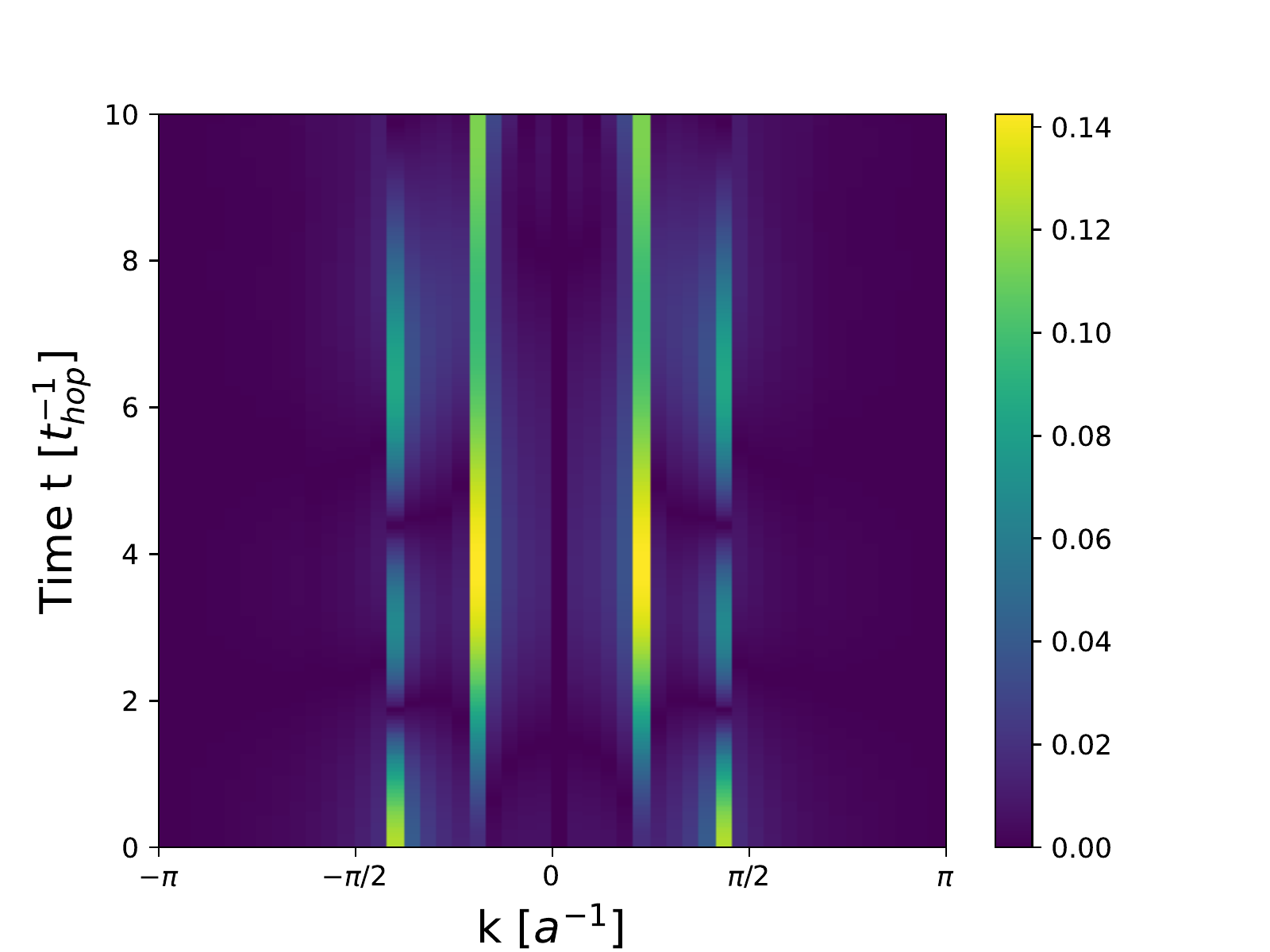}
\caption{tDMRG results for the time evolution of the local density after a quench in the $t$-$J_\perp$ model~\eqref{eq:tJperp} from $J_{\perp, {\rm initial}} = 1$ to $J_{\perp,{\rm final}} = 6$. (a) Time evolution as a function of position. (b) Fourier transform of the results of (a) to $k$-space by only considering the bulk region of (a).}
\label{fig:density}
\end{figure}
In Fig.~\ref{fig:density} we display a typical result for the time evolution of the local density following a quench starting from the gapless Luttinger liquid (LL) \cite{giamarchi} phase with dominant spin-density wave correlations \cite{PhysRevLett.107.115301,tJperp_PRA}. 
As can be seen, in the initial state typical Friedel-like oscillations are obtained~\cite{Bedurftig:1998p457,White:2002p348}, which are due to the open boundary conditions used.
The wave number of these oscillations is the Fermi momentum $k_F$ and is associated to the filling $n$.
For free spinful electrons, $k_F = k_F^\uparrow+ k_F^\downarrow$, where in the initial state $k_F^\uparrow=k_F^\downarrow=\pi n$.
The Fourier transform displayed in Fig.~\ref{fig:density}(b) shows that in the initial state the wave vector of the density oscillations is $k \approx 0.4 \pi$, in agreement with this expectation.
However, soon after the quench, an additional wave vector at $k \approx 0.2 \pi$ appears, which in the course of  time becomes dominant.
As this value is approximately half the 
original one, this insinuates that the Friedel-like density oscillations are now at half the Fermi momentum, as if one would have halfed the number of particles causing these density oscillations. 
Such behavior has been observed in the $t$-$J$ model in equilibrium~\cite{Moreno}, where the wave vector of the Friedel oscillations when increasing $J/t$ smoothly goes to half the original value upon entering the singlet-superconducting phase at low fillings~\cite{Moreno}, and it is also known from spin systems, where the doubling of the period can be associated to the formation of pairs of magnons \cite{BLBQ}. 
Taking on this picture, the quench seems to induce similar behavior, leading to the coexistence of two wave vectors in the oscillations of the local density.

\begin{figure}[t]
\includegraphics[width=\columnwidth]{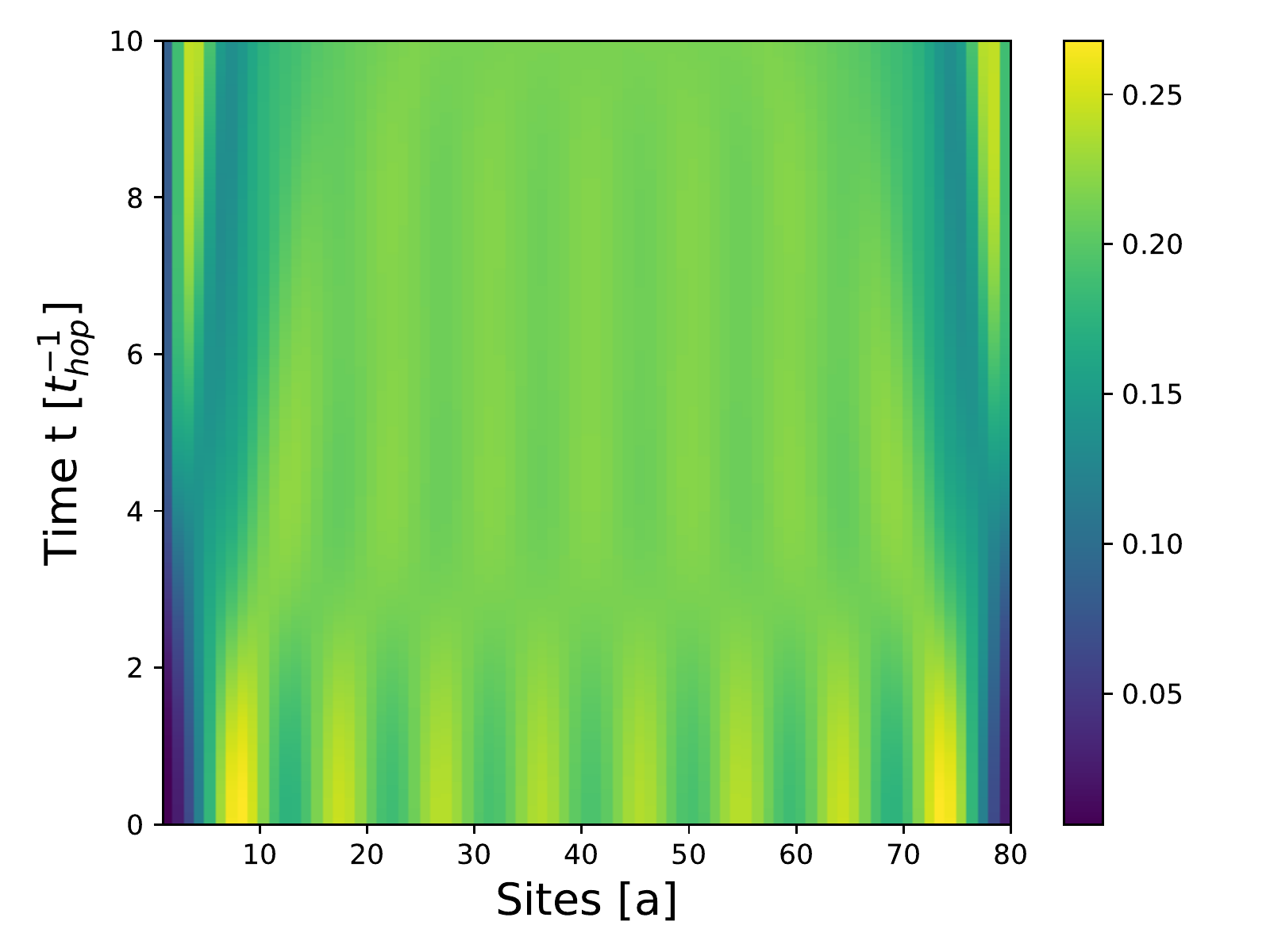}
\includegraphics[width=\columnwidth]{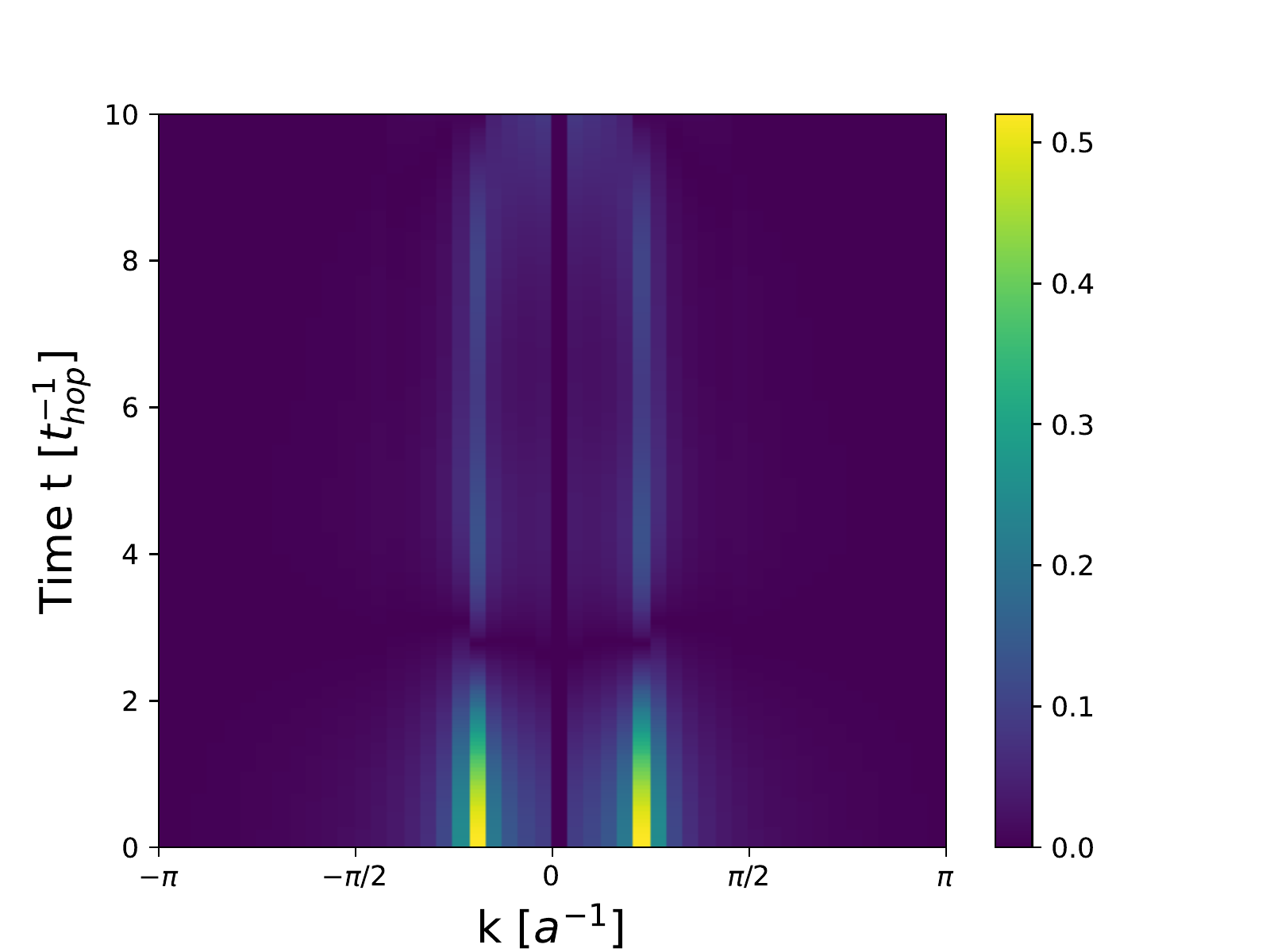}
\caption{tDMRG results for the time evolution of the local density after a quench in the $t$-$J_\perp$ model~\eqref{eq:tJperp} from $J_{\perp, {\rm initial}} = 5$ to $J_{\perp,{\rm final}} = 2$. (a) Time evolution as a function of position. (b) Fourier transform of the results of (a) to $k$-space by only considering the bulk region of (a). 
}
\label{fig:density2}
\end{figure}

When starting from the SC phase, instead, already in the initial state the wave vector of the Friedel-like oscillations is half the one of the LL phase at low $J_\perp/t_{\rm hop}$. 
The question arises, if the reverse effect might be realized, and a second wave vector with twice the value is obtained. 
However, this is not the case: as shown in Fig.~\ref{fig:density2}, in a quench from the SC phase to a smaller value of $J_\perp/t_{\rm hop}$ the density oscillations instead are strongly suppressed in the bulk of the system.

\subsection{Correlation functions}

\begin{figure*}[t]
\includegraphics[width=\columnwidth]{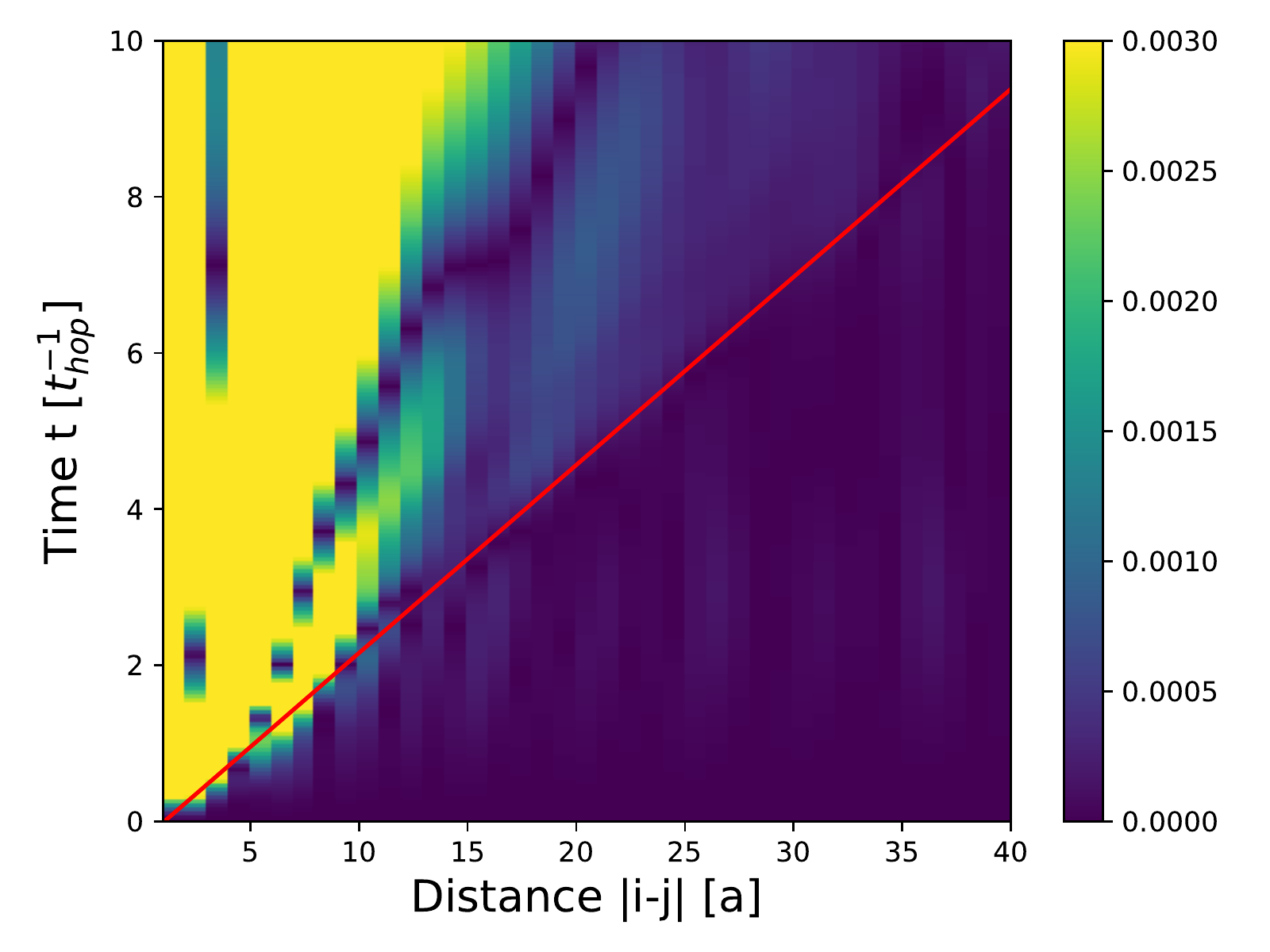}
\includegraphics[width=\columnwidth]{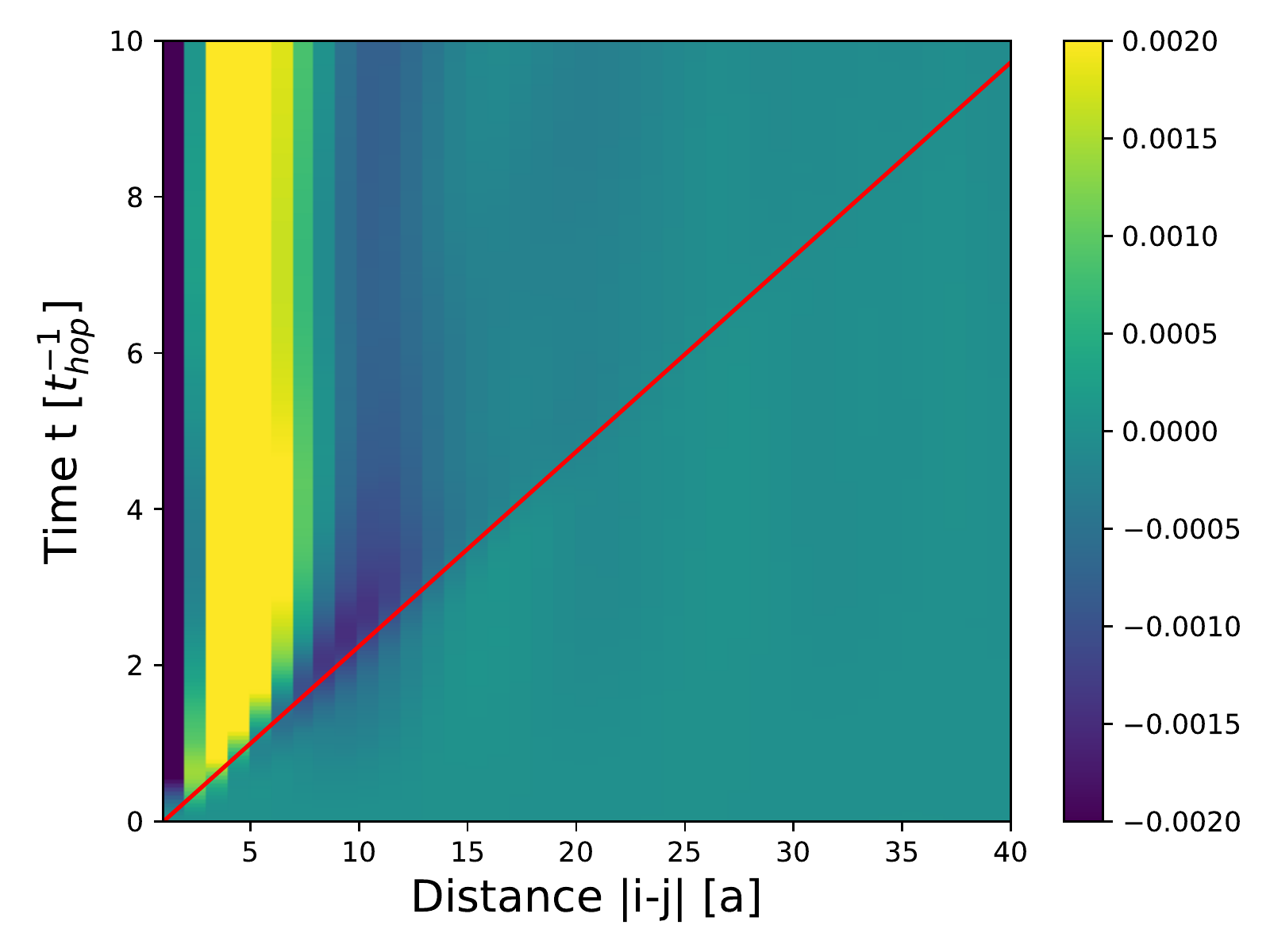}\\
\includegraphics[width=\columnwidth]{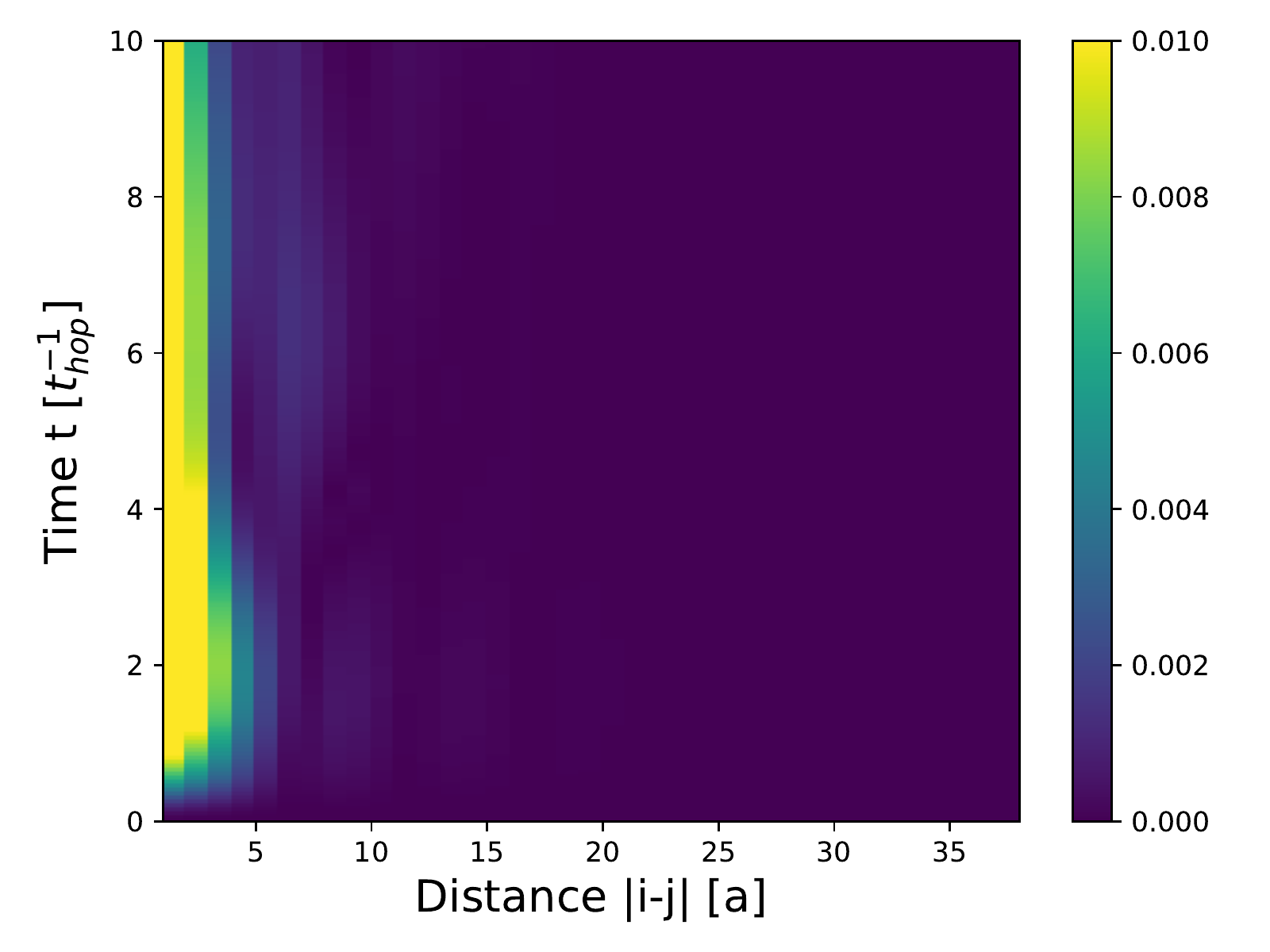}
\includegraphics[width=\columnwidth]{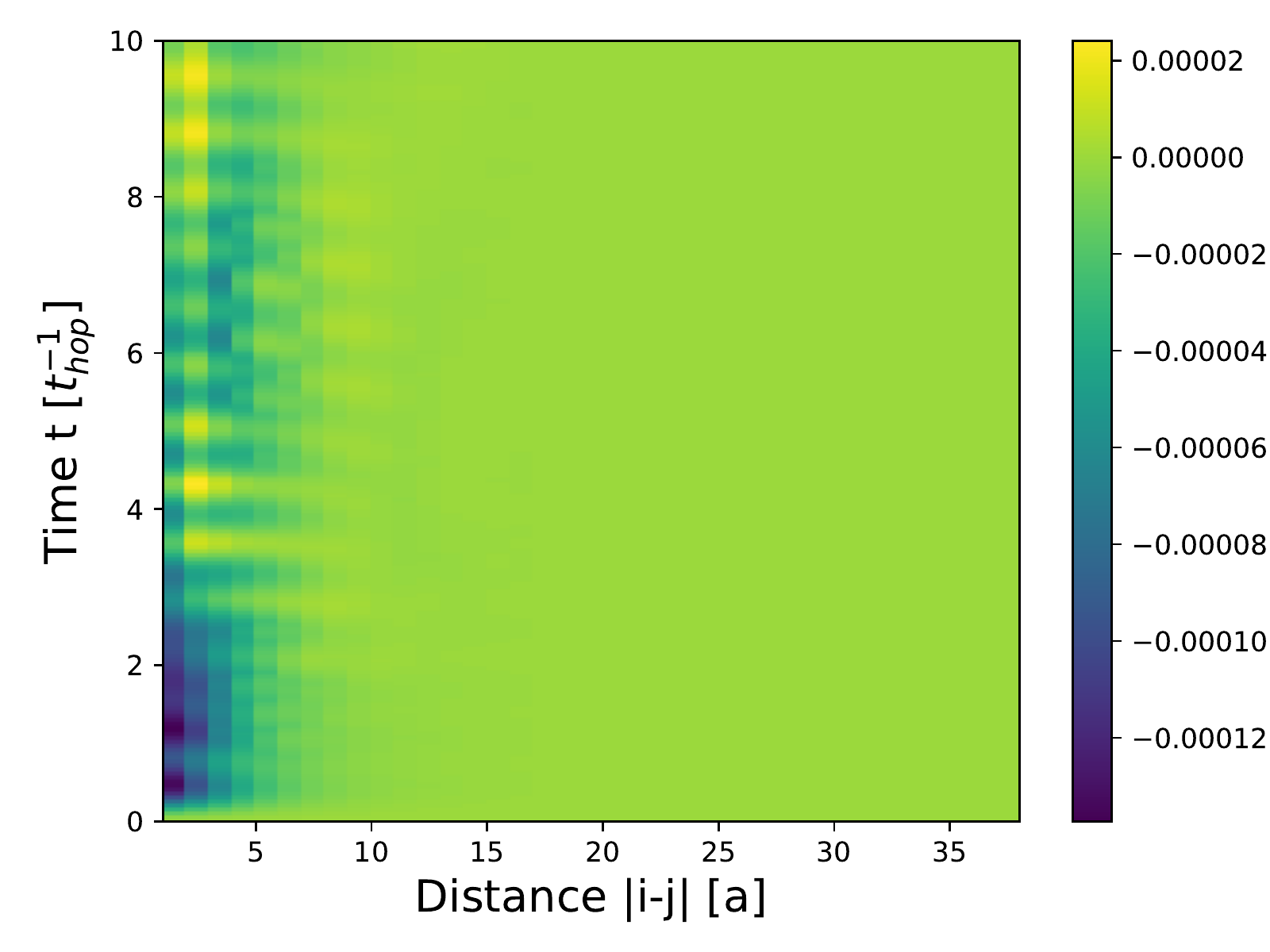}
\caption
{
tDMRG results for the time evolution of correlation functions following a quench in the $t$-$J_\perp$ model~\eqref{eq:tJperp} from $J_{\perp, {\rm initial}} = 1$ to $J_{\perp,{\rm final}} = 6$.
(a) Connected density-density correlation function~\eqref{eq:densdens}. (b) Connected longitudinal spin-spin correlation function~\eqref{eq:SzSz} (c) Singlet pairing and (d) Triplet pairing correlation functions according to Eqs.~\eqref{eq:pairing} -~\eqref{eq:delta_triplet}.
}
\label{fig:lightcone}
\end{figure*}

In Fig.~\ref{fig:lightcone} we contrast the time evolution of $N_{ij}$ and $C_{ij}^{\rm spin, long}$ 
to the one obtained for $P_{ij}^{T,S}$ for a quench from $J_\perp = 1$ to $J_\perp = 6$ at filling $n=0.2$. 
As can be seen, a clear light-cone signal is visible in $N_{ij}$ and $C_{ij}^{\rm spin, long}$, and similarly also in $C_{ij}^{\rm spin, trans}$ (not shown). 
The slope of the light-cone for quenches starting from $J_\perp = 1$ to values $J_\perp = 2, \, 3, \, 6$ is comparable and lies between 3 and 4.
Hence, the velocities of the quasi-particles, which one can associate to the light-cone, are comparable to the ones of free electrons, where the slope of the light-cone has the value four \cite{manmana09}.
This is also seen for quenches starting from $J_\perp = 5$ to the same final values. 
Furthermore, the slope of the light-cones for charge correlation functions and for spin correlation functions has comparable values.
This indicates that the quasi-particles involved in the formation of light-cones in these observables are not related to pairs of fermions, for which one could expect a slower velocity, as the mass of such an object would be twice the mass of a free particle.

Next, we ask if the same light-cone behavior is realized in pairing correlation functions.
Figure~\ref{fig:lightcone} contrasts the time evolution of the singlet and triplet pairing correlation functions to the ones of the density and spin correlation functions for the quench from $J_\perp = 1$ to $J_\perp = 6$.
Interestingly, the results do not show a clear linear signal at all. 
Instead, at short distances, a structure emerges, which then seems to freeze and not to move out to further distances.
In the course of time, oscillations are seen, which appear more pronounced in the triplet pairing correlation function. 

These results indicate that in both pairing correlation functions the light-cone signal is strongly suppressed.
Similar behavior is found when quenching from a SC phase to the LL phase at small $J_\perp$.
However, in this case, a faint linear signal can be obtained. 
Again, its slope is $\sim 4$. 
This indicates that the light-cone in principle can also be realized in the pairing correlation functions, but is substantially weaker than in the other correlation functions, an issue we will address in more detail in Sec.~\ref{sec:BCS}. 

The data presented in Fig.~\ref{fig:lightcone} (a) allow also the observation of a pattern in the dynamics inside the correlated region. There, it is possible to see at least another ballistic signal identified by a local maximum spreading with a velocity slower than $V_{lc}$. 
This is reminiscent of findings for exactly solvable models~\cite{Cevolani2017}, where internal patterns in the dynamics have been predicted and connected to phase velocities of the spectrum, in contrast to the light-cone given by the group velocity of excitations.

\section{Suppression of the light-cone in pairing correlation functions in BCS theory}
\label{sec:BCS}

The findings of the previous section may be specific to the $t$-$J$-model and to one spatial dimension.
In order to test their validity beyond this set-up, we now turn to a simple, analytically tractable model, which contains superconductivity. 
A simple approach is to recall BCS theory of superconductivity 
and to compute the time evolution of correlation functions after a quench in this framework. 
This leads us to the simple toy Hamiltonian
\begin{align}
\nonumber \mathcal{H} = &  \sum_{k} \left( \epsilon^{\phantom{\dagger}}_k -\mu \right) \left( c^\dagger_{k\uparrow} c^{\phantom{\dagger}}_{k\uparrow} + c^\dagger_{k\downarrow} c^{\phantom{\dagger}}_{k\downarrow} \right)- \\ & - \sum_k \left( \Delta \, c^\dagger_{k\uparrow} c^\dagger_{-k\downarrow} + \Delta^\ast \, c^{\phantom{\dagger}}_{-k\downarrow}c^{\phantom{\dagger}}_{k\uparrow}\right) \, ,
\label{eq:BCS}
\end{align}
where $\epsilon_k=-2\cos\left(k\right)$ is the dispersion of non-interacting electrons on a one-dimensional lattice, the operators $c^{(\dagger)}_{k\sigma}$ denote fermionic annihilation (creation) operators for a particle with momentum $k$ and spin $\sigma$ (note that the restriction of no double occupancy of the $t$-$J$-model is not valid here), and the gap $\Delta$ is assumed to be  independent of momentum $k$.
This toy Hamiltonian can be diagonalized using a Bogoliubov transformation
\begin{eqnarray}
 c_{k\uparrow}^\dagger &=& u_k \gamma_{k\uparrow}^\dagger +v_k^\ast \gamma_{-k \downarrow} \nonumber \\
 c_{k\uparrow} &=& u_k^\ast \gamma_{k\uparrow}+v_k \gamma_{-k\downarrow}^\dagger \nonumber \\
 c_{-k \downarrow} &=& u_k^\ast \gamma_{-k\downarrow} - v_k \gamma_{k\uparrow}^\dagger \nonumber \\
 c_{-k \downarrow}^\dagger  &=& u_k \gamma_{-k \downarrow}^\dagger -v_k^\ast \gamma_{k\uparrow} \, .
 \label{eq:bogoljubov}
\end{eqnarray}
The time evolution can then be computed using the following differential equations:
\begin{equation}
\imath \partial_t \begin{pmatrix} u_k(t) \\ v_k(t) \end{pmatrix} = \begin{pmatrix} \xi_k & \Delta_{f} \\ \Delta^{\ast}_f & -\xi_k \end{pmatrix} \begin{pmatrix} u_k(t) \\ v_k(t) \end{pmatrix} \, ,
\label{eq:EOM}
\end{equation}
where $\Delta_f\left( t \right)=\Delta_{eq}/\cosh \left( \Delta_{eq} t \right)$ solves the self-consistency equation as presented in Ref.~\onlinecite{Barankov2004}. 
For a quench from free fermions $\Delta_{eq} = 0$ to a value of $\Delta_{eq}=1/2$ the equations of motion~\eqref{eq:EOM} are integrated numerically to obtain the dynamics.
This allows us to 
compute the observables $N_{ij} (t)$, $P_{ij}^T (t)$, and $P^S_{ij}(t)$ defined in Eqs.~\eqref{eq:densdens} and \eqref{eq:pairing} -- \eqref{eq:delta_triplet}. 
Going from $\Delta=0$ to a finite value opens a gap in the spectrum of the excitations.
The computation of the analytical expressions for the time evolution of these observables using the BCS Hamiltonian is straight forward and the results are reproduced in App.~\ref{app:BCSCalc}.

\begin{figure}[h!]
\includegraphics[width=\columnwidth]{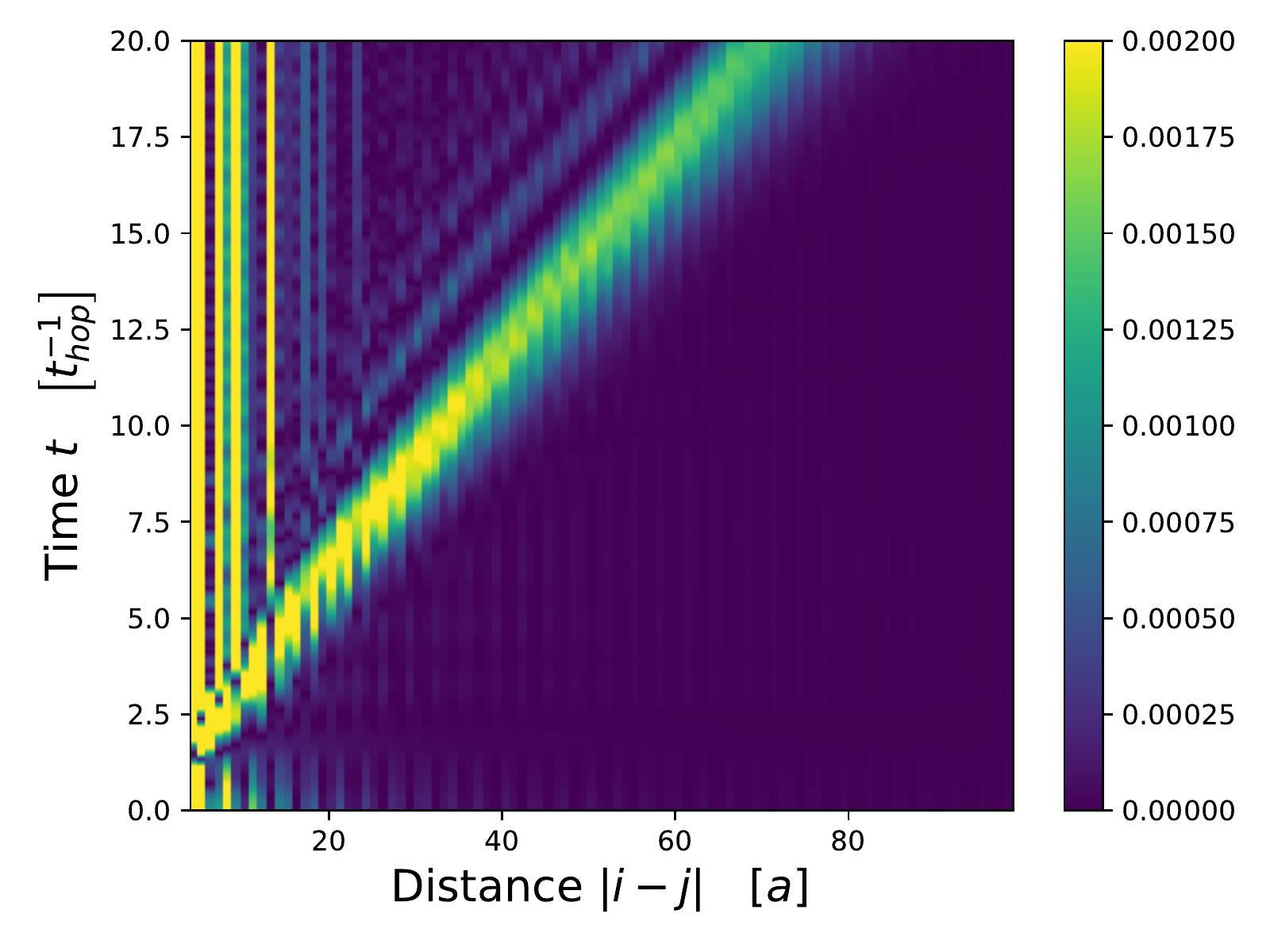}
\caption{\label{fig:dendenbcssc} Time evolution of the absolute value of the connected density-density correlation function Eq.~\eqref{eq:densdens} after a quench in the BCS Hamiltonian~\eqref{eq:BCS} from $\Delta_i = 0 \rightarrow \Delta_f = 1/2$.
The colormap has been set to enhance the contrast in the vicinity of the light-cone, where the maximal value is $\sim 0.002$.}
\end{figure}

We expect that if the behaviour of the different correlation functions obtained for the $t$-$J$ model is typical for superconductivity, we will also observe it in this simple BCS model. 
In Fig.~\ref{fig:dendenbcssc} we present the time evolution of the connected density-density correlation function Eq.~\eqref{eq:densdens} under the Hamiltonian~\eqref{eq:BCS}. 
The time evolution exhibits a linearly increasing correlation edge, which can be identified as the light-cone. 
Since the system cannot be described by quasi-particles with a well described spectrum, Eq.~\eqref{eq:spectrum}, we have no well defined maximum group velocity, but nevertheless the linear signal persists. 
In contrast, for the case where the time dependence of the gap can be neglected, we have a well defined excitation spectrum that can be used to predict the velocity of the light-cone with great accuracy, see App.~\ref{app:BCSCalc}.

\begin{figure}[h!]
\includegraphics[width=\columnwidth]{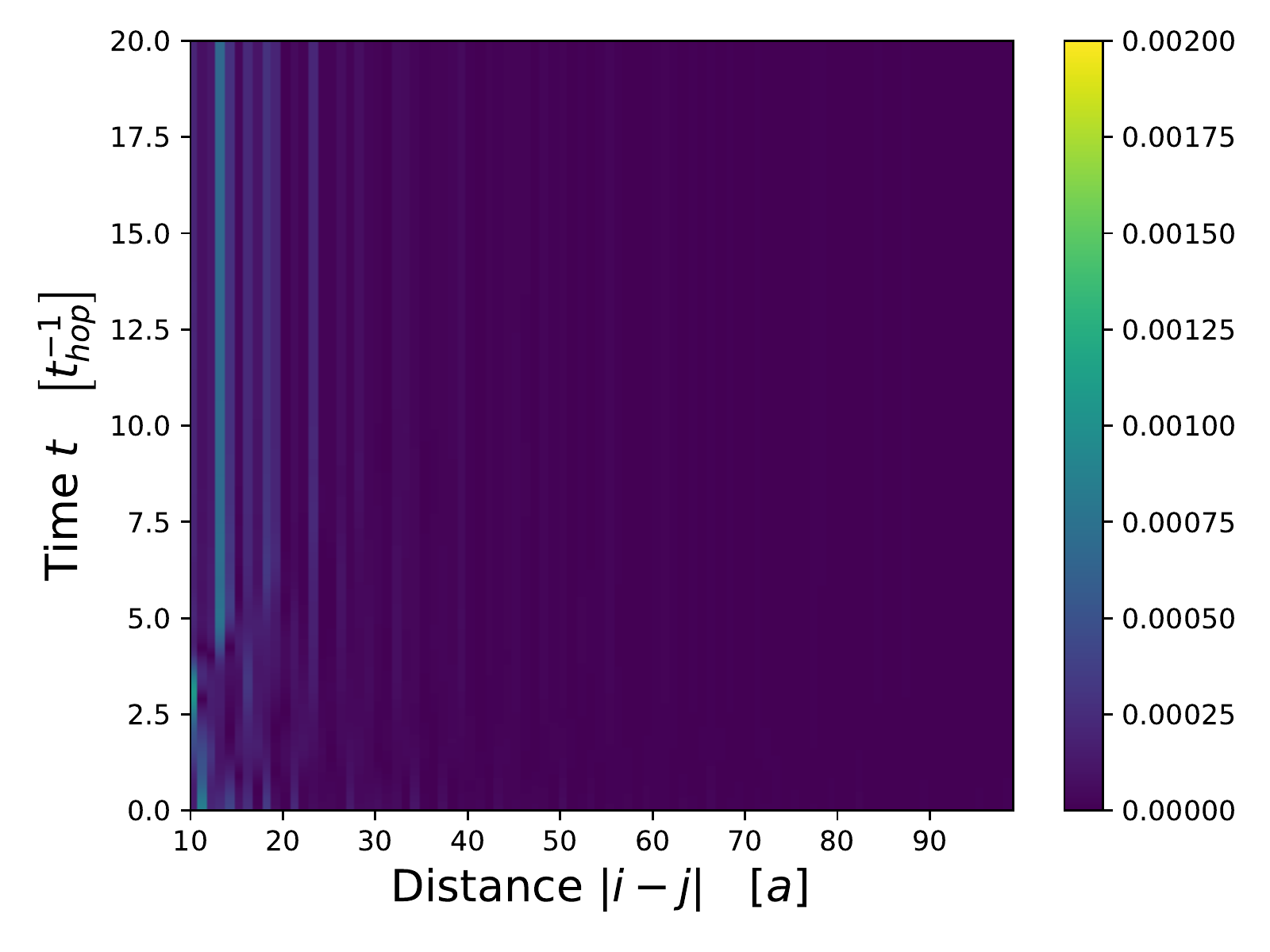}
\caption{\label{fig:tripletBCSsc} Absolute value of the connected triplet correlation function Eq.~\eqref{eq:triplet} as function of distance and time after a quench in the BCS Hamiltonian~\eqref{eq:BCS} from $\Delta_i = 0 \rightarrow \Delta_f = 1/2$. 
The colormap has been set to be equal to the one used in Fig.~\ref{fig:dendenbcssc}.} 
\end{figure}

\begin{figure}[h!]
\includegraphics[width=\columnwidth]{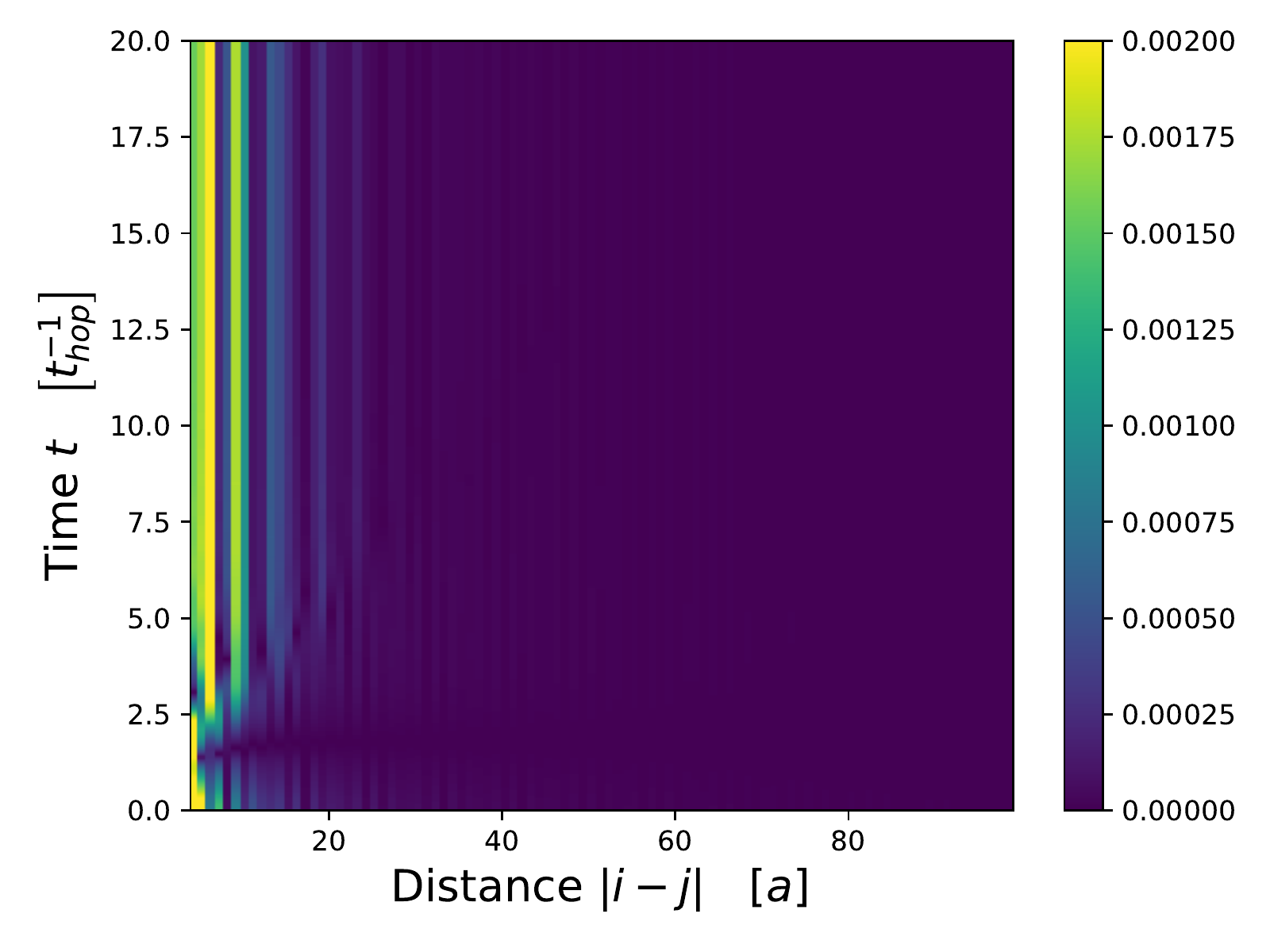}
\caption{\label{fig:singletBCSsc}
 Absolute value of the connected singlet correlation function Eq.~\eqref{eq:singletpairbcsappendix} as function of distance and time after a quench in the BCS Hamiltonian~\eqref{eq:BCS} from $\Delta_i = 0 \rightarrow \Delta_f = 1/2$. The colormap has been set to be equal as the one used in Fig.~\ref{fig:dendenbcssc} in order to see how no traveling signal is present in this figure.
}
\end{figure}

In Figs.~\ref{fig:tripletBCSsc} and~\ref{fig:singletBCSsc} we plot the time evolution of the singlet and triplet correlation functions as defined in Eqs.~\eqref{eq:pairing}--\eqref{eq:delta_triplet} (since in the BCS treatment the expectation values for $\langle \Delta^{(\dagger)}(i)_{T,S}\rangle$ can be finite, we consider the connected correlation functions). 
Hence, similar to the tDMRG results for the $t$-$J$ model, also in this self-consistent BCS treatment the light-cone is strongly suppressed or absent in the pairing correlation functions.
Note that at smaller distances, there is a larger signal, which does not move towards further distances, similar to the findings in the $t$-$J$ model.\\

Hence, also in BCS theory we see a light-cone in certain correlation functions, but in the pairing correlation functions it is strongly suppressed.
This is true for both, a time-dependent gap equation as well as when neglecting the time-dependence of the gap.
This behavior can be further analyzed in the simple quadratic fermionic model studied in App.~\ref{sec:appendix_BCS}.
As discussed in more detail there, one obtains that in the mean-field theory the expressions for the density-density correlation functions and the pairing correlation functions differ by terms, which can be summarized as 
\begin{equation}\label{eq:niunjd}
\langle n_{i\downarrow}n_{j\uparrow}\rangle +\langle n_{i\uparrow}n_{j\downarrow}\rangle \, .
\end{equation}
In the mean-field approach, however, it is this term, which carries most of the weight of the light-cone signal.
As this contribution is absent in the pairing correlation functions, the light-cone signal is suppressed there. 
This is true also for two-dimensional systems, as discussed in App.~\ref{app:2dmean}.



\section{Conclusions}
\label{sec:conclusions}
We considered the time evolution of local observables and different correlation functions following a quantum quench of the $t$-$J_\perp$ chain Eq.~\eqref{eq:tJperp} at fillings $n = 0.2$.
For quenches from a gapless initial state to a spin-gapped singlet-superconducting phase, the original Friedel oscillations change their wave vector from $k_F$ to $k_F/2$ at short times, and then a coexistence of both wave vectors is seen.
Instead, for quenches in the reverse direction, the Friedel oscillations in the initial state are with wave vector $k_F/2$, and then are suppressed in the course of the time evolution.\\
In the correlation functions one clearly observes light-cone behavior in the density-density and spin correlation functions, but a strongly suppressed light-cone in the pairing correlation functions.
The slope of the light-cone is approximately the same in all cases.

Similar behavior is obtained in quenches in a BCS treatment.
In one dimension, the values of the pairing correlation functions in the vicinity of the light-cone are strongly suppressed compared to the ones of the density-density correlation function.
This can be understood by the lack of contribution of the term~\eqref{eq:niunjd} in the pairing correlation functions, which carries most of the weight of the density correlation function in the vicinity of the light-cone. 

The superconducting phases in the one-dimensional systems treated here are not true SC phases according to the Mermin-Wagner-Hohenberg theorem. 
While a microscopic treatment of the time evolution of $t$-$J$ models in two dimensions is presently out of reach, we applied the BCS treatment also to two-dimensional systems, where similar behavior was identified.
It would be interesting to investigate this in higher dimensions using more refined techniques.

\section*{Acknowledgements} 
We thank Markus Schmitt for useful discussions, in particular for pointing out the possibility to do a BCS treatment.
Financial support from the Deutsche Forschungsgemeinschaft (DFG) through SFB/CRC1073 (Project B03) is gratefully acknowledged.

\appendix
\section{Analytical expressions for the time evolution of the correlation functions in time-dependent BCS theory}\label{app:BCSCalc}
\label{sec:appendix_BCS}

We study the BCS-Hamiltonian~\eqref{eq:BCS} using the Bogoljubov-transformation~\eqref{eq:bogoljubov}. In Sec.~\ref{sec:BCS} we discussed the dynamics in the presence of a dynamical gap $\Delta_f$, here we want to discuss a more simplistic picture where we have just a constant gap $\Delta_f$.
In the set of equations~\eqref{eq:bogoljubov}, $u_k$ and $v_k$ are complex parameters that can be chosen to diagonalize the Hamiltonian~\eqref{eq:BCS} using $\xi_k=\epsilon_k-\mu$, 
\begin{align}
&  \vert u_k \vert^2= \frac{1}{2}\left( 1+\frac{\xi_k}{\sqrt{\xi_k + \vert \Delta_f \vert^2}} \right) \\
& \vert v_k \vert^2 = \frac{1}{2}\left( 1- \frac{\xi_k}{\sqrt{\xi_k + \vert \Delta_f \vert^2}} \right) \, .
\end{align}
To study the time evolution, we promote the amplitudes $u_k$ and $v_k$ to be time dependent complex variables.
Their explicit form is then determined solving the Heisenberg equation of motion, see Ref.~\cite{Barankov2004, Natu2013}
\begin{equation}
\imath \partial_t \begin{pmatrix} u_k(t) \\ v_k(t) \end{pmatrix} = \begin{pmatrix} \xi_k & \Delta_{f} \\ \Delta^{\ast}_f & -\xi_k \end{pmatrix} \begin{pmatrix} u_k(t) \\ v_k(t) \end{pmatrix} \, .
\end{equation}
The initial conditions are then given by $u_k(t=0)=1$ and $v_k(t=0)=0$. 
This leads to 
\begin{align*}
& u_k(t) = \cos\left( E_k^f t \right)-\imath \frac{\xi_k}{E_k^f}\sin\left( E_k^f t \right) \\
& v_k(t) = -\imath \frac{\Delta_f^\ast}{E_k^f} \sin \left( E_k^f t \right) \, .
\end{align*}
The spectrum 
\begin{equation} 
E_{k\uparrow} = E_{k\downarrow} = \sqrt{\left(\epsilon_k -\mu\right)^2 + \vert \Delta \vert^2}
\label{eq:spectrum}
\end{equation} 
is gapless for $\Delta=0$ and gapped otherwise. For the specific quench we will study in the followings, $\Delta_i=0\rightarrow \Delta_f=1/2$, it is possible to extract the maximum velocity of $3.1$ which is consistent with the data found in the tDMRG.\\
These results can be used to compute the time evolution of expectation values of different observables.
We find it useful to introduce the following functions, in terms of which the time evolution of more complicated observables can be expressed: 
\begin{align*}
& \mathcal{U}_{l\uparrow} \left(R,t\right)= \frac{1}{L}\sum_k e^{\imath k R} \vert u_k \vert^2 n_{k\uparrow}^0 \\
& \mathcal{U}_{u\uparrow} \left(R,t\right)= \frac{1}{L}\sum_k e^{\imath k R} \vert u_k \vert^2 \left(1-n_{k\uparrow}^0\right) \\
& \mathcal{U}_{l\downarrow} \left(R,t\right)= \frac{1}{L}\sum_k e^{\imath k R} \vert u_k \vert^2 n_{k\downarrow}^0 \\
& \mathcal{U}_{u\downarrow} \left(R,t\right)= \frac{1}{L}\sum_k e^{\imath k R} \vert u_k \vert^2 \left(1-n_{k\downarrow}^0\right) \\
& \mathcal{V}_{l\uparrow} \left(R,t\right)= \frac{1}{L}\sum_k e^{\imath k R} \vert v_k \vert^2 n_{k\uparrow}^0 \\
& \mathcal{V}_{u\uparrow} \left(R,t\right)= \frac{1}{L}\sum_k e^{\imath k R} \vert v_k \vert^2 \left(1-n_{k\uparrow}^0\right) \\
& \mathcal{V}_{l\downarrow} \left(R,t\right)= \frac{1}{L}\sum_k e^{\imath k R} \vert v_k \vert^2 n_{k\downarrow}^0 \\
& \mathcal{V}_{u\downarrow} \left(R,t\right)= \frac{1}{L}\sum_k e^{\imath k R} \vert v_k \vert^2 \left(1-n_{k\downarrow}^0\right) \\
& \mathcal{UV}\left( R,t \right) = \frac{1}{L}\sum_{k} e^{\imath k R} u_k v_k^\ast \left( 1-n_{k\uparrow}^0 -n_{k\downarrow}^0 \right) \, ,
\end{align*}
where all of them depend on $R=\vert i-j \vert$ because the system has translational invariance.
If our initial state is invariant under a spin flip, $n_{k\uparrow}=n_{k\downarrow}$, then the subscripts $\uparrow$ and $\downarrow$ can be suppressed. 
We hence consider in the following only one type per function. 

\subsection{Density-density correlations}

\begin{figure}[h!]
\includegraphics[width=\columnwidth]{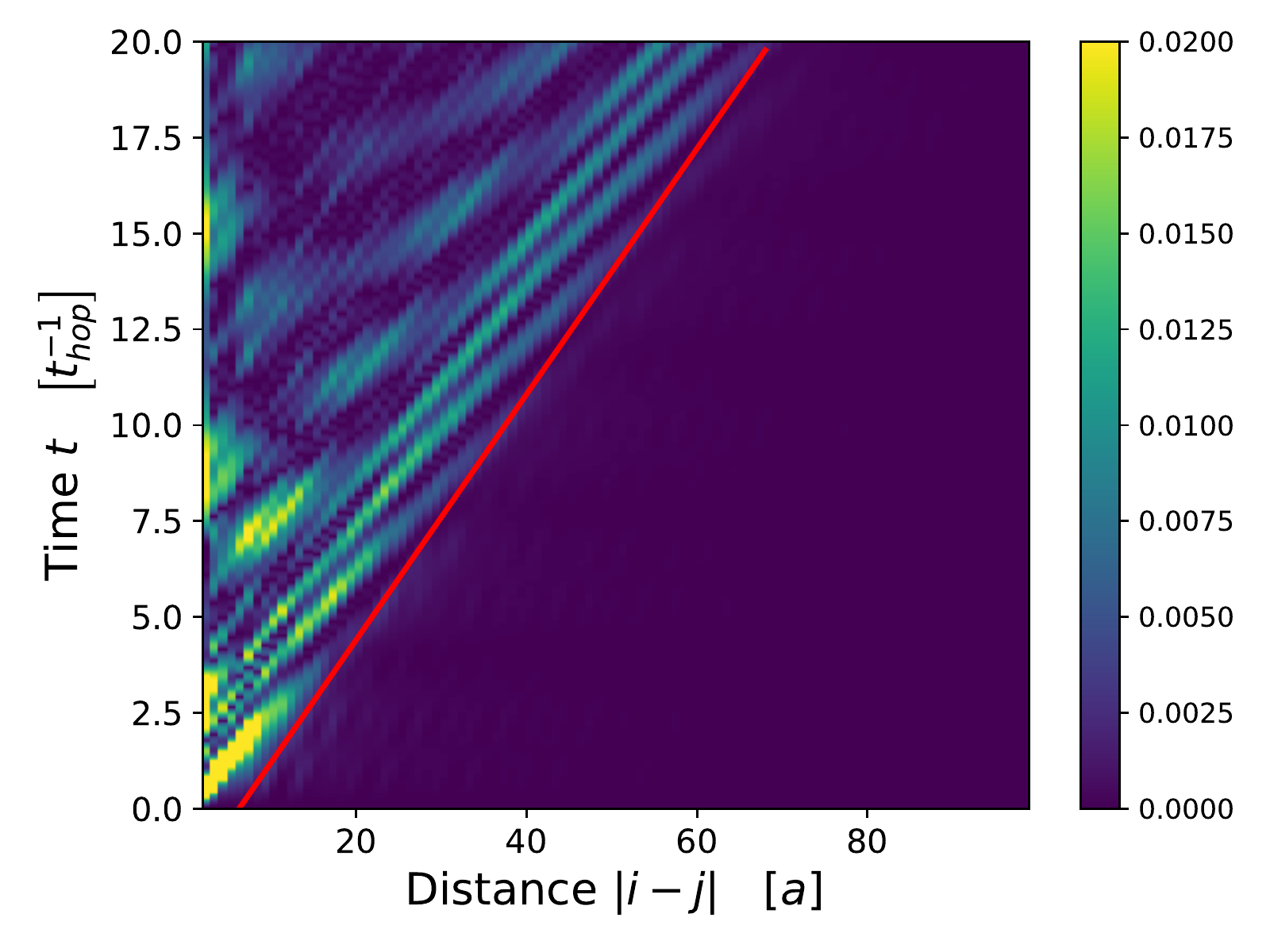}
\caption{\label{fig:dendenbcs} Time evolution of the absolute value of the connected density-density correlation function Eq.~\eqref{eq:densdens} after a quench in the BCS Hamiltonian~\eqref{eq:BCS} from $\Delta_i = 0 \rightarrow \Delta_f = 1/2$. The red solid line has the slope of  the maximal group velocity of the spectrum Eq.~\eqref{eq:spectrum} multiplied by two and agrees very well with the border of the light-cone region. 
The colormap has been set to enhance the contrast in the vicinity of the light-cone, where the maximal value is $\sim 0.01$.}
\end{figure}

In this section, we compute the time evolution of the correlator Eq.~\eqref{eq:densdens}. 
In time-dependent BCS theory, we can rewrite this observable using the expressions of the previous section and obtain
\begin{align}
\mathcal{C}\left(R,t\right) = & \left(\mathcal{U}_l +\mathcal{V}_u \right)\left(\mathcal{U}_u^\ast +\mathcal{V}_l^\ast \right) + \left(\mathcal{U}_l^\ast +\mathcal{V}_u^\ast \right)\left(\mathcal{U}_u +\mathcal{V}_l \right) \nonumber \\
& + 2 \vert \mathcal{UV} \vert^2 \, ,
\label{eq:densdensbcsappendix}
\end{align}
where the functions $\mathcal{U}$ and $\mathcal{V}$ depend on the distance $R$ and on time $t$.
We then consider a quench $\Delta_i=0\rightarrow \Delta_f=1/2$ for a system at zero chemical potential $\mu = 0$, with initial conditions $n_{k\uparrow}=n_{k\downarrow}=\theta\left( k_F - \vert k \vert \right)$, and filling 0.2, i.e., $k_F=\frac{2\pi}{5}$.
The result is displayed in Fig.~\ref{fig:dendenbcs}. 

\subsection{Triplet-pairing correlation function}

We can now study the connected triplet-pairing correlation function, which we define as
\begin{align}
& P^T \left(R,t\right)  = \left\langle \Delta^{\dagger}_T(i) \Delta^{\phantom{\dagger}}_T(j)  \right\rangle  - \left\langle \Delta^{\dagger}_T(i) \right\rangle \left\langle \Delta^{\phantom{\dagger}}_T(j) \right\rangle \\
& \mbox{with } \Delta^{\phantom{\dagger}}_T(i) =c^{\phantom{\dagger}}_{i\uparrow} c^{\phantom{\dagger}}_{i+1\,\uparrow} \, .
\end{align}
Using the same simplifications used for the density-density correlations, we obtain
\begin{align}
& P^T_{R=\vert i-j\vert }  =    \mathcal{F} \left(R-1, t \right) \mathcal{F} \left(R+1, t \right)- \mathcal{F}^2 \left(R, t \right) \, , \label{eq:triplet}\\
& \mbox{where } \mathcal{F} \left(R, t \right) = \left[ \mathcal{U}_l \left( R,t \right)+\mathcal{V}_u \left(R,t\right) \right] \, .
\end{align}
This result shows that the time evolution of the triplet-pairing correlation function is obtained by subtracting the product of the same function at neighboring lattice points $R$, $R-1$, and $R+1$.
This explains why its value is so small, as the total contribution is in fact a quartic contribution in the according Taylor expansion.

\begin{figure}[h!]
\includegraphics[width=\columnwidth]{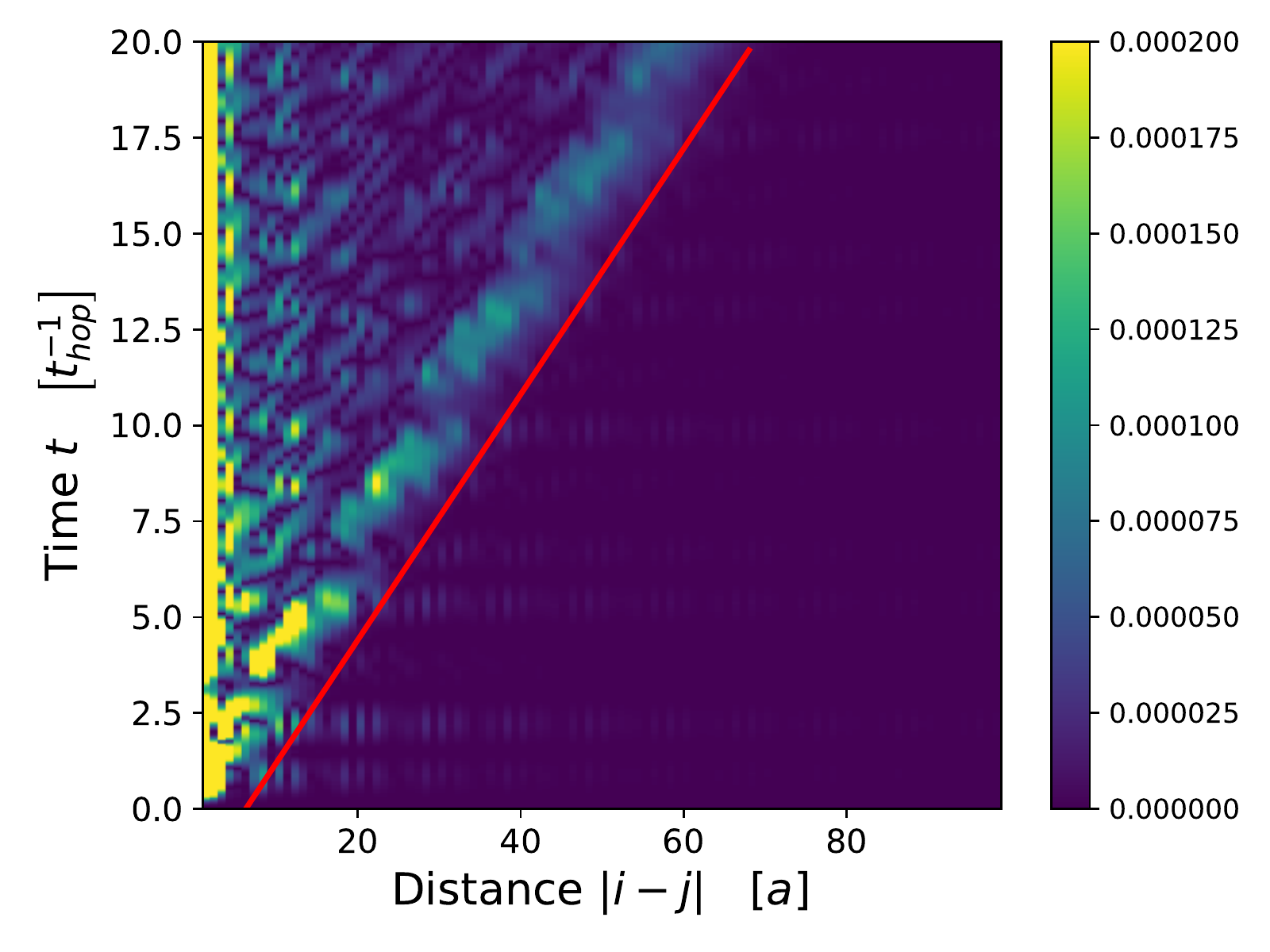}
\caption{\label{fig:tripletBCS} Absolute value of the connected triplet correlation function Eq.~\eqref{eq:triplet} as function of distance and time after a quench in the BCS Hamiltonian~\eqref{eq:BCS} from $\Delta_i = 0 \rightarrow \Delta_f = 1/2$. The red solid line has the slope of  the maximal group velocity of the spectrum Eq.~\eqref{eq:spectrum} multiplied by two and agrees very well with the border of the light-cone region. 
The colormap has been set to enhance the contrast in the vicinity of the light-cone, where the maximal value is $\sim 0.0001$. Note that close to the light-cone, the strength of the signal is two orders of magnitude smaller than the one of the density-density correlation function in Fig.~\ref{fig:dendenbcs}.}
\end{figure}

\subsection{Singlet-pairing correlation function}

Finally we can write down the time evolution of the connected singlet-pairing correlation function

\begin{align}
& P^S \left(R,t\right)  = \left\langle \Delta^{\dagger}_S(i) \Delta^{\phantom{\dagger}}_S(j)  \right\rangle  - \left\langle \Delta^{\dagger}_S(i) \right\rangle \left\langle \Delta^{\phantom{\dagger}}_S(j) \right\rangle \\
& \mbox{with } \Delta^{\dagger}_S(i) =
\frac{1}{\sqrt{2}} \left(c^\dagger_{i,\downarrow} c^\dagger_{i+1,\uparrow} - c^\dagger_{i,\uparrow} c^\dagger_{i+1,\downarrow} \right) \, .
\end{align}
As before, the exact expression can be written as function of the building blocks presented before, leading to
\begin{equation}
\label{eq:singletpairbcsappendix}
P^S \left( R,t \right) =  \mathcal{F} \left(R-1, t \right) \mathcal{F} \left(R+1, t \right)+ \mathcal{F}^2 \left(R, t \right)
\end{equation}
where $\mathcal{F}$ has been introduced in the previous section. 
In this case there is no subtraction of terms that could explain why this correlation function is so small compared to the density-density one. 
However, a closer look at the different fundamental building blocks involved in the expressions~\eqref{eq:densdensbcsappendix} and~\eqref{eq:singletpairbcsappendix} shows that the difference is indeed the expression~\eqref{eq:niunjd}. As this is found to carry most of the weight in the density-density correlation function, the values of the singlet-pairing correlation functions are also strongly suppressed.

\begin{figure}[h!]
\includegraphics[width=\columnwidth]{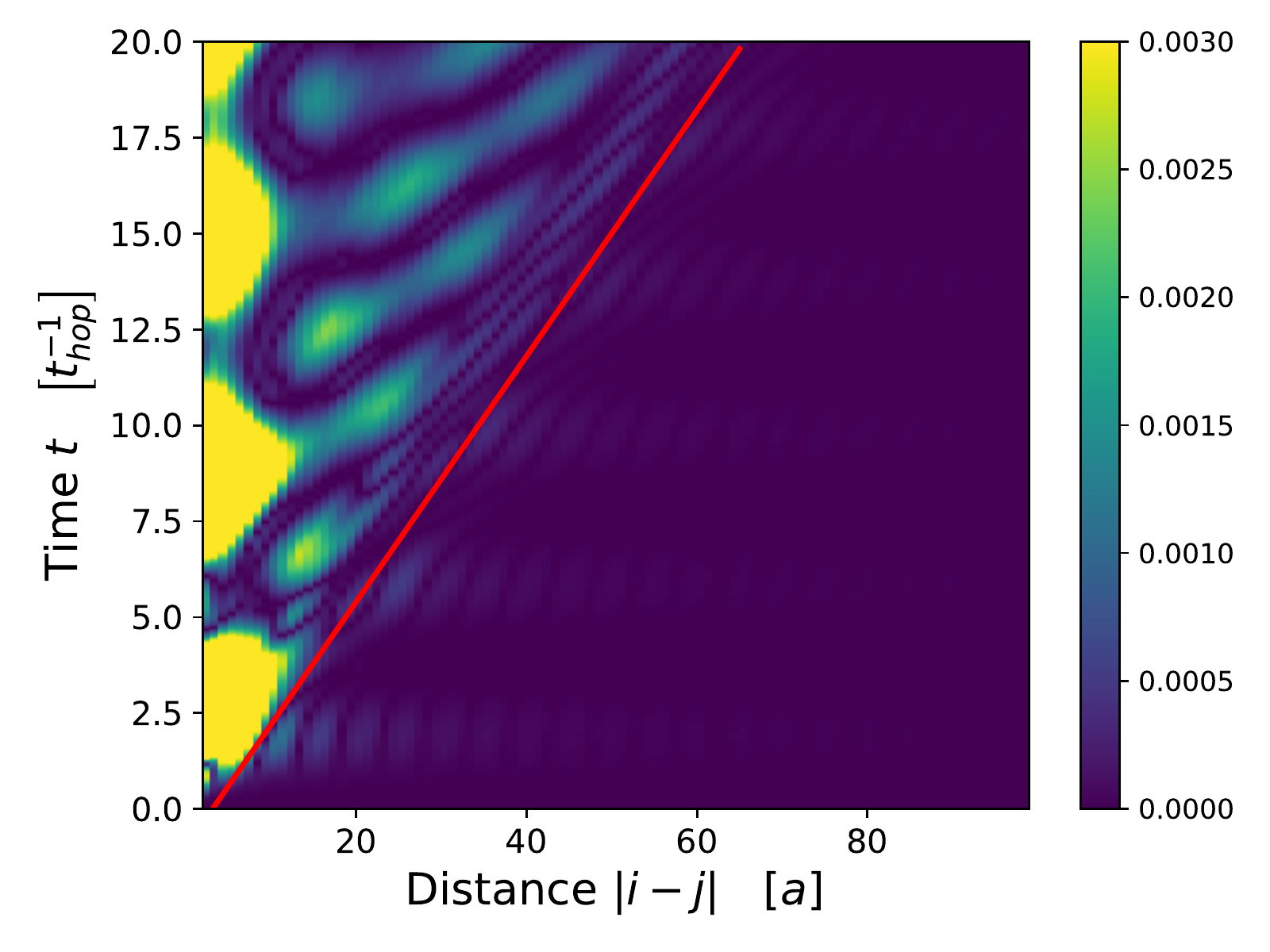}
\caption{\label{fig:singletBCS}
 Absolute value of the connected singlet correlation function Eq.~\eqref{eq:singletpairbcsappendix} as function of distance and time after a quench in the BCS Hamiltonian~\eqref{eq:BCS} from $\Delta_i = 0 \rightarrow \Delta_f = 1/2$. The red solid line has the slope of  the maximal group velocity of the spectrum Eq.~\eqref{eq:spectrum} multiplied by two and agrees very well with the border of the light-cone region. 
The colormap has been set to enhance the contrast in the vicinity of the light-cone, where the maximal value is $\sim 0.001$. Note that close to the light-cone, the strength of the signal is between one and two orders of magnitude smaller than the one of the density-density correlation function in Fig.~\ref{fig:dendenbcs}.
}
\end{figure}

\subsection{Relevant correlations in the free model}
\begin{figure}[h!]
\includegraphics[width=\columnwidth]{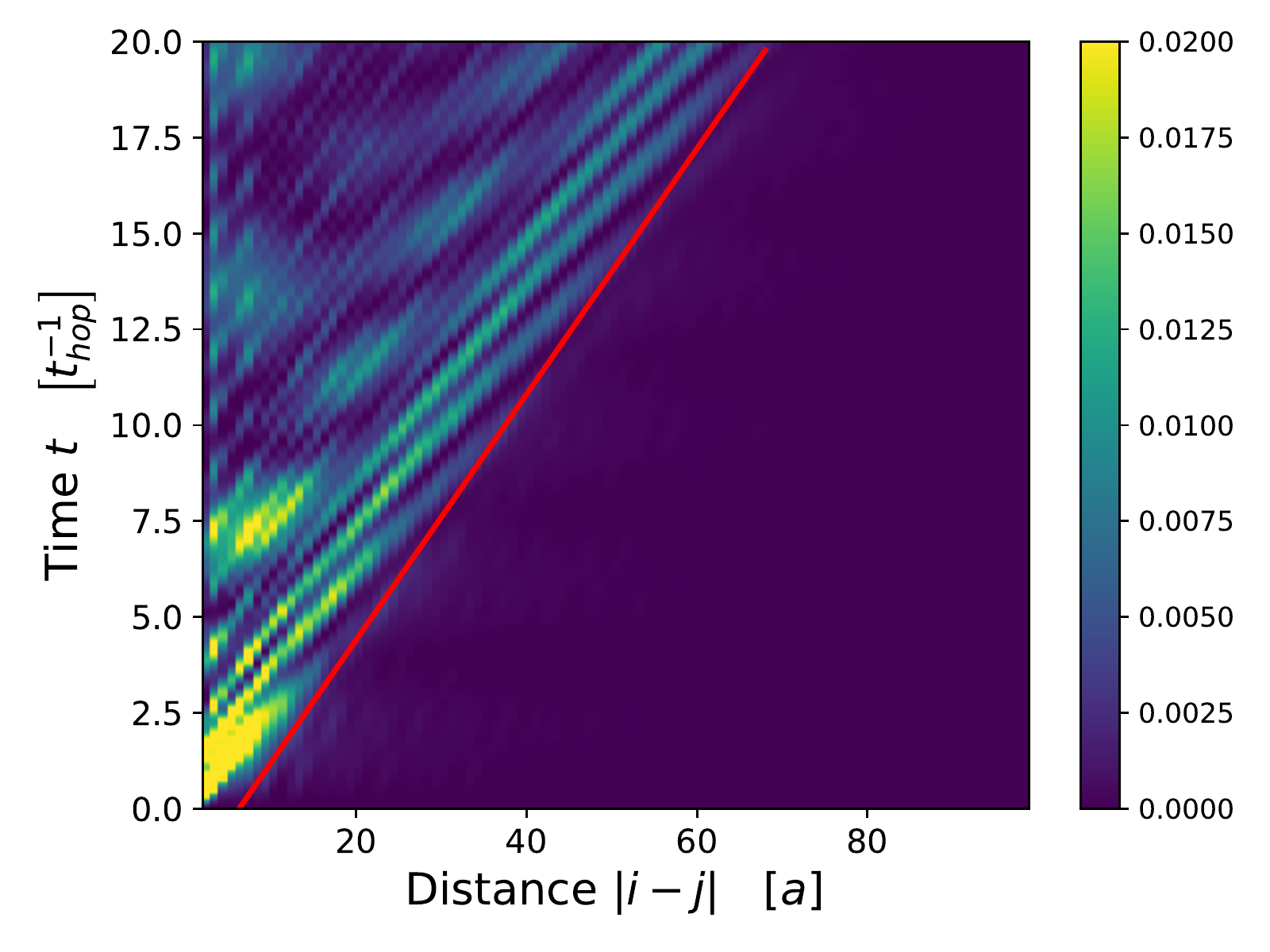}
\caption{\label{fig:BCSSpinOp}Absolute value of the dynamics of the observable Eq.~\eqref{eq:niunjd} as function of distance and time after a quench in the BCS Hamiltonian~\eqref{eq:BCS} from $\Delta_i = 0 \rightarrow \Delta_f = 1/2$. The red solid line has the slope of  the maximal group velocity of the spectrum Eq.~\eqref{eq:spectrum} multiplied by two and agrees very well with the border of the light-cone region. 
The colormap has been set to enhance the contrast in the vicinity of the light-cone, where the maximal value is $\sim 0.01$, which is comparable to the one of the density-density correlation function in Fig.~\ref{fig:dendenbcs}.
}
\end{figure}

A closer look at the results for $\mathcal{C}\left(R,t\right)$ \eqref{eq:densdensbcsappendix} shows that the main contribution to the strong signal in the light-cone in the density-density correlation function is due to the term 
\begin{equation}\label{eq:appnni}
\langle n_{i\downarrow}n_{j\uparrow}\rangle +\langle n_{i\uparrow}n_{j\downarrow}\rangle.
\end{equation}
The dynamics of this observable alone is plotted in Fig.~\ref{fig:BCSSpinOp}. 
If we compare it to the density-density correlations, Fig.~\ref{fig:dendenbcs}, the difference of the values between the two in the vicinity of the light-cone is extremely small.
From Eq.~\eqref{eq:densdensbcsappendix}~\eqref{eq:triplet}~\eqref{eq:singletpairbcsappendix} one obtains that the term~\eqref{eq:appnni} does not contribute to the pairing correlation functions, which explains why their values in the vicinity of the light-cone are substantially smaller.\\
We can now compare these data with the ones obtained in Sec.~\ref{sec:BCS} using self-consistency. In the free model, a signal is present in all the different observables. For the density-density correlations a strong signal is present and it is around one order of magnitude larger compared to the one obtained with the self-consistency condition. Regarding the pair correlation in the free model a light-cone signal is present in the dynamics, albeit at least two orders of magnitude smaller than the one in the corresponding density-density correlation. The same observable computed using self-consistency show a visibly inhibited signal, which is close to the numerical results from the tDMRG.

\section{Time-dependent BCS approach in two dimensions}\label{app:2dmean}
\begin{figure}
\includegraphics[width=\columnwidth]{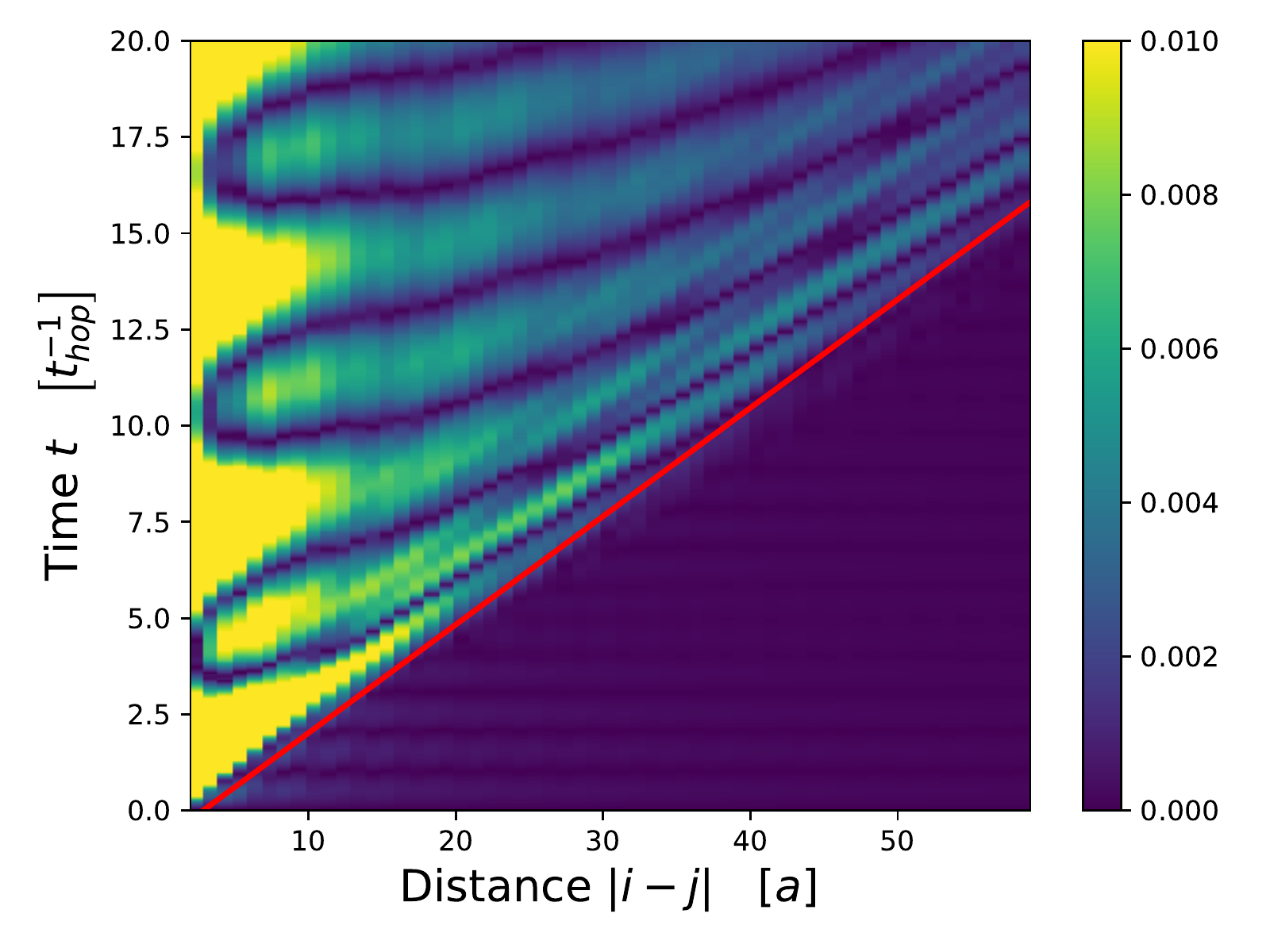}
\caption{\label{fig:denden2D} Time evolution of the absolute value of the connected density-density correlation function Eq.~\eqref{eq:densdensbcsappendix} after a quench in the BCS Hamiltonian~\eqref{eq:BCS} from $\Delta_i = 0 \rightarrow \Delta_f = 1/2$ in a two-dimensional system. The coordinate $R$ labels the points on the line $\left( R,R \right)$, bisectrix of the plane. The red solid line has the slope of  the maximal group velocity of the spectrum Eq.~\eqref{eq:spectrum} multiplied by two and agrees very well with the border of the light-cone region. 
The colormap has been set to enhance the contrast in the vicinity of the light-cone, where the maximal value is $\sim 0.008$.}
\end{figure}

\begin{figure}
\includegraphics[width=\columnwidth]{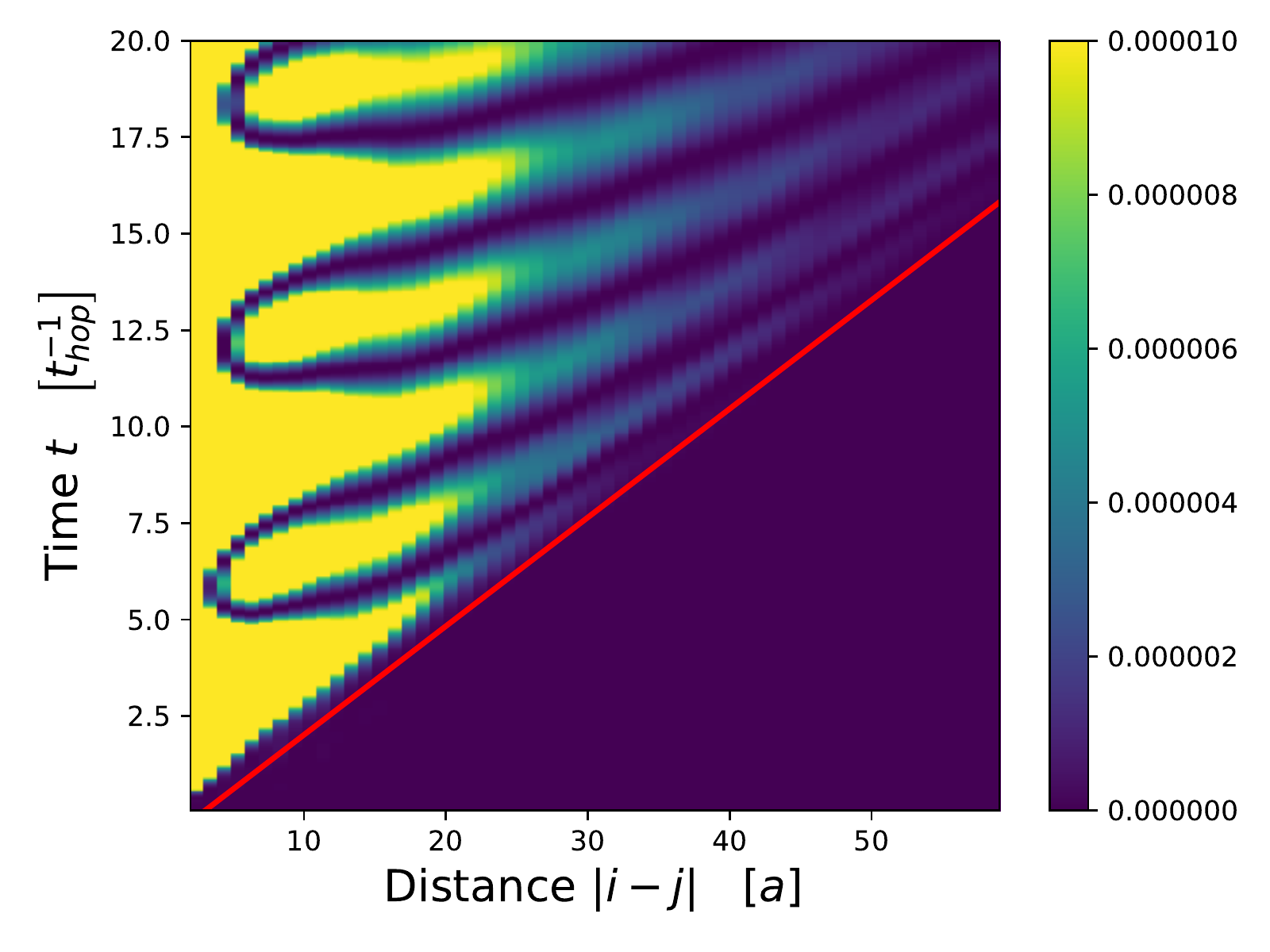}
\caption{\label{fig:triplet2D} Absolute value of the on-site pairing correlation function Eq.~\eqref{eq:sebastian} as function of distance and time after a quench in the BCS Hamiltonian~\eqref{eq:BCS} from $\Delta_i = 0 \rightarrow \Delta_f = 1/2$ in a two-dimensional system. The coordinate $R$ labels the points on the line $\left(R,R \right)$, bisectrix of the plane. The red solid line has the slope of  the maximal group velocity of the spectrum Eq.~\eqref{eq:spectrum} multiplied by two and agrees very well with the border of the light-cone region. 
The colormap has been set to enhance the contrast in the vicinity of the light-cone, where the maximal value is $\sim 8 \cdot 10^{-6}$. 
Note that close to the light-cone, the strength of the signal is three orders of magnitude smaller than the one of the density-density correlation function in Fig.~\ref{fig:denden2D}.}
\end{figure}

The toy model of Hamiltonian~\eqref{eq:BCS} can easily be generalized to two-dimensional situations. 
The spatial indices become vectors, $i=\left( i_x,i_y \right)$, and correspondingly the Fourier vector $k=\left( k_x,k_y \right)$. 
The equation of motion for the amplitudes $u_k$ and $v_k$ takes exactly the same form and it can be solved analytically again.\\
For the quench $\Delta_{\textrm{i}}=0\rightarrow \Delta_{\textrm{f}}=1/2$ we study the density-density correlation function, which takes the same general expression of Eq.~\eqref{eq:densdensbcsappendix} where $R$ labels the two dimensional distance.\\
Because of the $2D$ geometry, the possibilities to define pair correlation functions are richer. 
Here, for the sake of simplicity, we consider only \textit{on-site} pairing correlation functions defined as:
\begin{align}
 \mathcal{S}\left( R, t \right) & =\langle c_{i \uparrow}^\dagger \left( t \right) c_{i\downarrow}^\dagger \left( t \right) c_{j \downarrow} \left( t \right) c_{j \uparrow} \left( t \right) \rangle = \vert \mathcal{F}\left( R, t \right) \vert^2\, ,   
 \label{eq:sebastian}
\end{align}
which correspond to $s$-wave pairing. 
For $D=1$, the light-cone in $\mathcal{S}$ has the same suppression as the correlators $P^T$ and $P^S$ compared to the density-density correlation function.\\
In Fig.~\ref{fig:denden2D} we plot the time evolution of the connected density-density correlation function along the line $\left(R,R\right)$ as function of time. 
The time evolution of $\mathcal{S}\left(R,t\right)$ along the same line is plotted in Fig.~\ref{fig:triplet2D}. 
As in the 1D case, we observe a suppression of the amplitude of the light-cone from the density-density to the \textit{on-site} correlation function.

\begin{figure}
\includegraphics[width=\columnwidth]{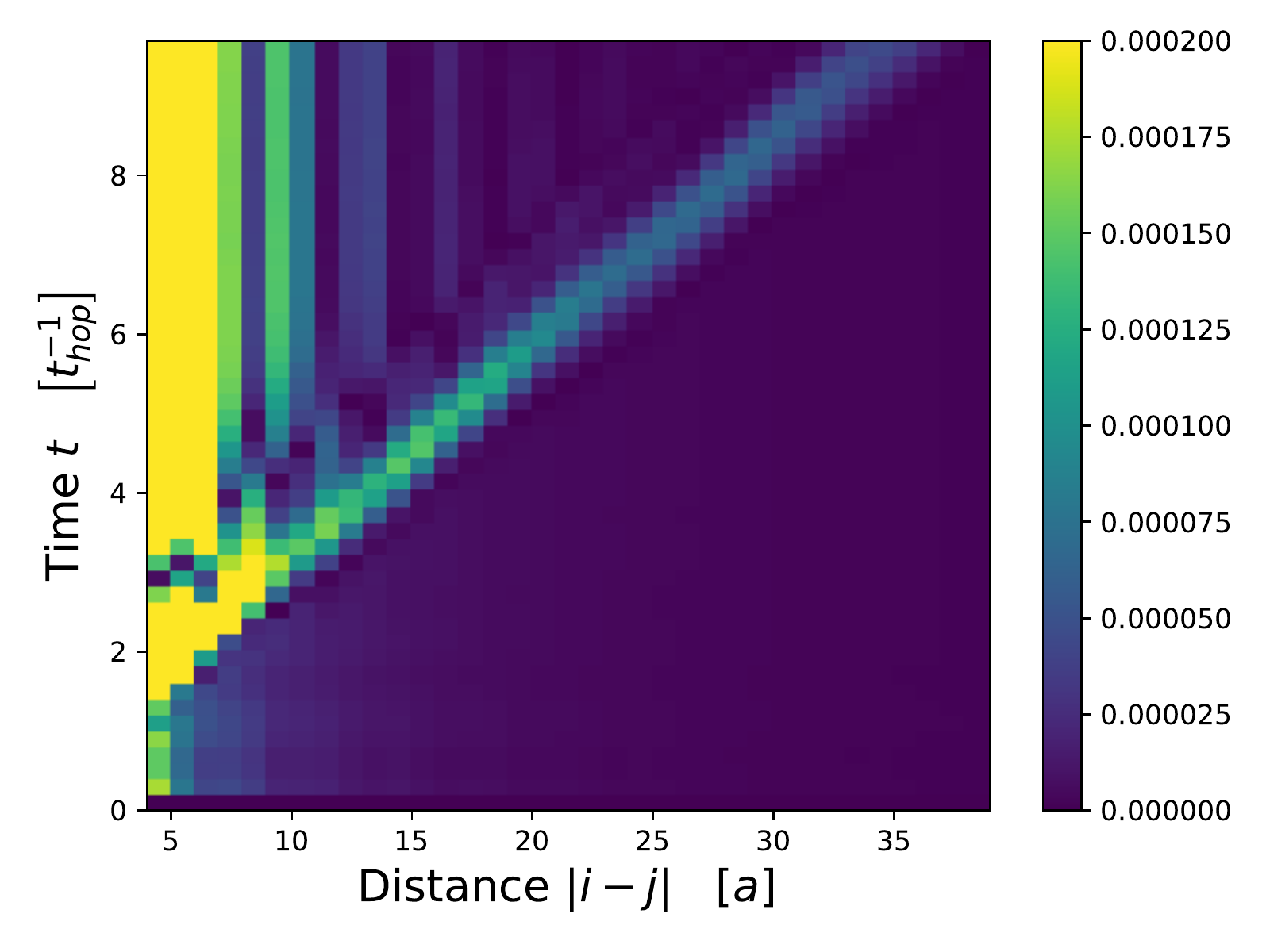}
\caption{\label{fig:selfcondenden2D} Time evolution of the absolute value of the connected density-density correlation function Eq.~\eqref{eq:densdensbcsappendix} after a quench in the BCS Hamiltonian~\eqref{eq:BCS} from $\Delta_i = 0 \rightarrow \Delta_f = 1/2$ in a two-dimensional system. The time evolution is computed using the self-consistent BCS theory. The coordinate $R$ labels the points on the line $\left( R,R \right)$, bisectrix of the plane. The colormap has been set to enhance the light-cone signal.}
\end{figure}

\begin{figure}
\includegraphics[width=\columnwidth]{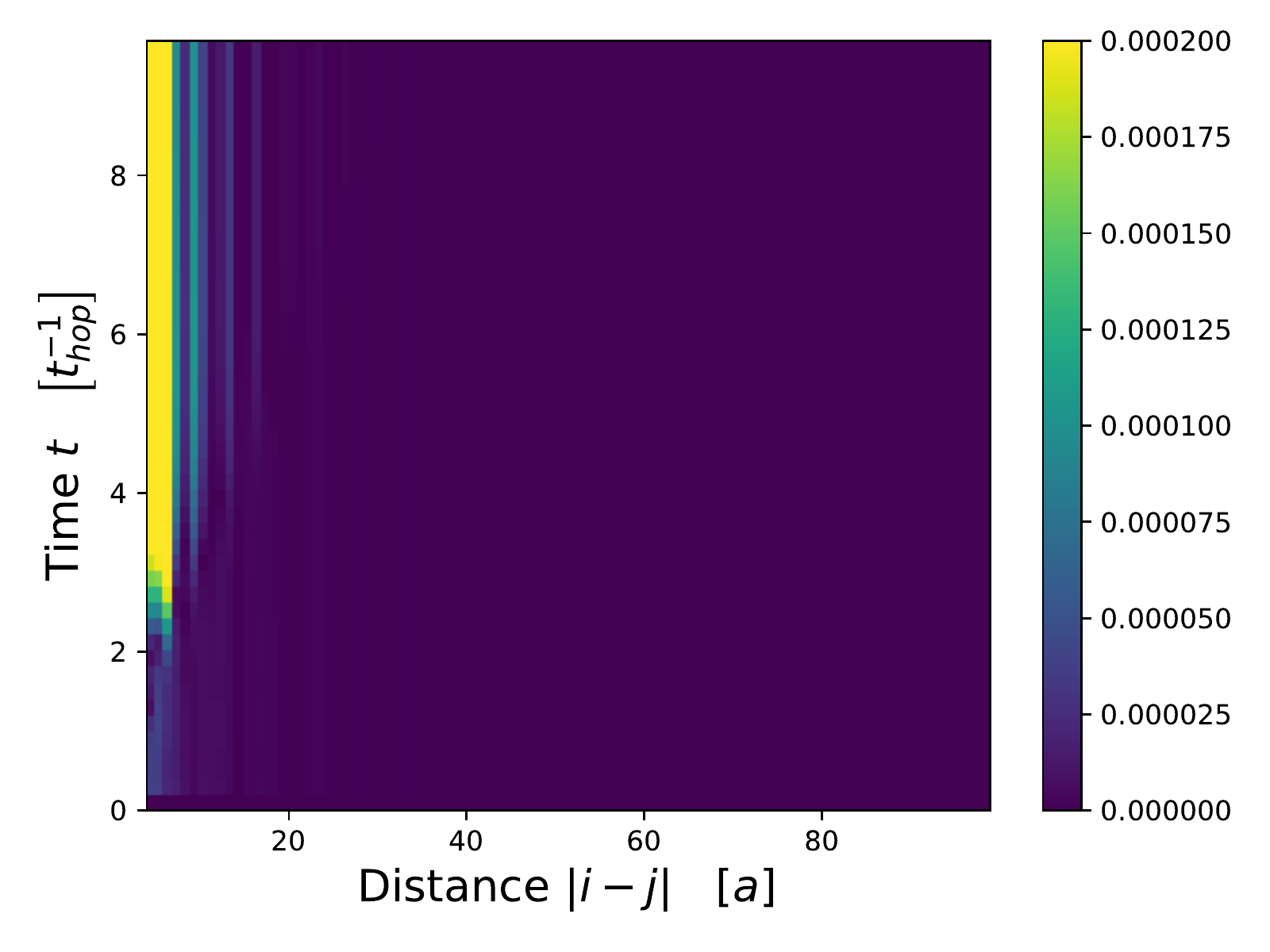}
\caption{\label{fig:selfcontriplet2D} Absolute value of the on-site pairing correlation function Eq.~\eqref{eq:sebastian} as function of distance and time after a quench in the BCS Hamiltonian~\eqref{eq:BCS} from $\Delta_i = 0 \rightarrow \Delta_f = 1/2$ in a two-dimensional system. The dynamics is computed using self-consistent BCS theory. The coordinate $R$ labels the points on the line $\left(R,R \right)$, bisectrix of the plane.
If we compare it to Fig.~\ref{fig:selfcondenden2D} we can see that we have an agreement with the $D=1$ case computed using self-consistent equations.}
\end{figure}

Similarly to the $D=1$, it is possible to study the dynamics using the self-consistent BCS theory.
In Fig.~\ref{fig:selfcondenden2D} the dynamics of the density-density correlations in a two dimensional system are presented. It is possible to clearly see a travelling signal. In Fig.~\ref{fig:selfcontriplet2D} the dynamics of on-site pairing correlations of a two dimensional system are presented. In this case it is not possible to see any ballistic signal in the time evolution. These results are in agreement with the ones presented in Sec.~\ref{sec:BCS} for the $D=1$ case.


\newpage

\begin{thebibliography}{10}

\bibitem{LiebRobinson72}
E.~H. Lieb and D.~W. Robinson, Commun. Math. Phys. {\bf 28},  251  (1972).

\bibitem{bravyi2006lieb}
S. Bravyi, M. Hastings, and F. Verstraete, Physical review letters {\bf 97},
  050401  (2006).

\bibitem{Kastner_NJP}
M. Kastner, New Journal of Physics {\bf 17},  123024  (2015).

\bibitem{calabrese06}
P. Calabrese and J. Cardy, Phys. Rev. Lett. {\bf 96},  136801  (2006).

\bibitem{calabrese07}
P. Calabrese and J. Cardy, J. Stat. Mech.: Theor. and Exp.  P06008  (2007).

\bibitem{laeuchli08}
A.~M. L\"auchli and C. Kollath, J. Stat. Mech.: Theor. and Exp.  P05018
  (2008).

\bibitem{manmana09}
S.~R. Manmana, S. Wessel, R.~M. Noack, and A. Muramatsu, Phys. Rev. B {\bf 79},
   155104  (2009).

\bibitem{Carleo2014}
G. Carleo, F. Becca, L. Sanchez-Palencia, S. Sorella, and M. Fabrizio, Phys.
  Rev. A {\bf 89},  031602  (2014).

\bibitem{Cheneau:2012bh}
M. Cheneau, P. Barmettler, D. Poletti, M. Endres, P. Schau{\ss}, T. Fukuhara,
  C. Gross, I. Bloch, C. Kollath, and S. Kuhr, Nature {\bf 481},  484  (2012).

\bibitem{lightcone_Langen2013}
T. Langen, R. Geiger, M. Kuhnert, B. Rauer, and J. Schmidmayer, Nature Physics
  {\bf 9},  640  (2013).

\bibitem{Hastings2006}
M.~B. Hastings and T. Koma, Communications in Mathematical Physics {\bf 265},
  781  (2006).

\bibitem{PRA_Correlations}
K.~R.~A. Hazzard, M. van~den Worm, M. Foss-Feig, S.~R. Manmana, E.~G.
  Dalla~Torre, T. Pfau, M. Kastner, and A.~M. Rey, Phys. Rev. A {\bf 90},
  063622  (2014).

\bibitem{PRL_Eisert}
J. Eisert, M. van~den Worm, S.~R. Manmana, and M. Kastner, Phys. Rev. Lett.
  {\bf 111},  260401  (2013).

\bibitem{PRL_ZheXuan}
Z.-X. Gong, M. Foss-Feig, S. Michalakis, and A.~V. Gorshkov, Phys. Rev. Lett.
  {\bf 113},  030602  (2014).

\bibitem{Hauke2013}
P. Hauke and L. Tagliacozzo, Phys. Rev. Lett. {\bf 111},  207202  (2013).

\bibitem{Cevolani2015}
L. Cevolani, G. Carleo, and L. Sanchez-Palencia, Phys. Rev. A {\bf 92},  041603
   (2015).

\bibitem{Cevolani2016}
L. Cevolani, G. Carleo, and L. Sanchez-Palencia, New Journal of Physics {\bf
  18},  093002  (2016).

\bibitem{Cevolani2017}
L. {Cevolani}, J. {Despres}, G. {Carleo}, L. {Tagliacozzo}, and L.
  {Sanchez-Palencia}, arXiv:1706.00838 (2017).

\bibitem{Frerot2018}
I. Fr\'erot, P. Naldesi, and T. Roscilde, Phys. Rev. Lett. {\bf 120},  050401
  (2018).

\bibitem{Vodola2014}
D. Vodola, L. Lepori, E. Ercolessi, A.~V. Gorshkov, and G. Pupillo, Phys. Rev.
  Lett. {\bf 113},  156402  (2014).

\bibitem{Foss-Feig2015}
M. Foss-Feig, Z.-X. Gong, C.~W. Clark, and A.~V. Gorshkov, Phys. Rev. Lett.
  {\bf 114},  157201  (2015).

\bibitem{Richerme2014}
P. Richerme, Z.-X. Gong, A. Lee, C. Senko, J. Smith, M. Foss-Feig, S.
  Michalakis, A.~V. Gorshkov, and C. Monroe, Nature {\bf 511},  198  (2014).

\bibitem{Schachenmayer2015}
J. Schachenmayer, A. Pikovski, and A.~M. Rey, New Journal of Physics {\bf 17},
  065009  (2015).

\bibitem{Buyskikh2016}
A.~S. Buyskikh, M. Fagotti, J. Schachenmayer, F. Essler, and A.~J. Daley, Phys.
  Rev. A {\bf 93},  053620  (2016).

\bibitem{Jurcevic2014}
P. Jurcevic, B.~P. Lanyon, P. Hauke, C. Hempel, P. Zoller, R. Blatt, and C.~F.
  Roos, Nature {\bf 511},  202  (2014).

\bibitem{Regemortel2016}
M. Van~Regemortel, D. Sels, and M. Wouters, Phys. Rev. A {\bf 93},  032311
  (2016).

\bibitem{PhysRevLett.107.115301}
A.~V. Gorshkov, S.~R. Manmana, G. Chen, J. Ye, E. Demler, M.~D. Lukin, and
  A.~M. Rey, Phys. Rev. Lett. {\bf 107},  115301  (2011).

\bibitem{giamarchi}
T. Giamarchi, {\em Quantum Physics in One Dimension}, Vol.~121 of {\em
  International Series of Monographs on Physics} (Oxford University Press,
  Oxford, 2004).

\bibitem{merminwagner_orig}
N.~D. Mermin and H. Wagner, Phys. Rev. Lett. {\bf 17},  1133  (1966).

\bibitem{merminwagner_erratum}
N.~D. Mermin and H. Wagner, Phys. Rev. Lett. {\bf 17},  1307  (1966).

\bibitem{hohenberg_orig}
P.~C. Hohenberg, Phys. Rev. {\bf 158},  383  (1967).

\bibitem{tJperp_PRA}
S.~R. Manmana, M. M\"oller, R. Gezzi, and K.~R.~A. Hazzard, Phys. Rev. A {\bf
  96},  043618  (2017).

\bibitem{reviewmolecules}
L.~D. Carr, D. DeMille, R.~V. Krems, and J. Ye, New Journal of Physics {\bf
  11},  055049  (2009).

\bibitem{Lemeshko_Review}
M. Lemeshko, R.~V. Krems, J.~M. Doyle, and S. Kais, Molecular Physics {\bf
  111},  1648  (2013).

\bibitem{PhysRevLett.101.133004}
J. Deiglmayr, A. Grochola, M. Repp, K. M\"ortlbauer, C. Gl\"uck, J. Lange, O.
  Dulieu, R. Wester, and M. Weidem\"uller, Phys. Rev. Lett. {\bf 101},  133004
  (2008).

\bibitem{Silke_science}
K.-K. Ni, S. Ospelkaus, M.~H.~G. de~Miranda, A. Pe'er, B. Neyenhuis, J.~J.
  Zirbel, S. Kotochigova, P.~S. Julienne, D.~S. Jin, and J. Ye, Science {\bf
  322},  231  (2008).

\bibitem{Silke_science2}
S. Ospelkaus, K.-K. Ni, D. Wang, M.~H.~G. de~Miranda, B. Neyenhuis, G.
  Qu\'em\'ener, P.~S. Julienne, J.~L. Bohn, D.~S. Jin, and J. Ye, Science {\bf
  327},  853  (2010).

\bibitem{Silke_Nature}
K.~K. Ni, S. Ospelkaus, D. Wang, G. Quemener, B. Neyenhuis, M.~H.~G.
  de~Miranda, J.~L. Bohn, J. Ye, and D.~S. Jin, Nature {\bf 464},  1324
  (2010).

\bibitem{PhysRevLett.104.030402}
S. Ospelkaus, K.-K. Ni, G. Qu\'em\'ener, B. Neyenhuis, D. Wang, M.~H.~G.
  de~Miranda, J.~L. Bohn, J. Ye, and D.~S. Jin, Phys. Rev. Lett. {\bf 104},
  030402  (2010).

\bibitem{Amodsen}
A. Chotia, B. Neyenhuis, S.~A. Moses, B. Yan, J.~P. Covey, M. Foss-Feig, A.~M.
  Rey, D.~S. Jin, and J. Ye, Phys. Rev. Lett. {\bf 108},  080405  (2012).

\bibitem{PhysRevLett.84.246}
A.~N. Nikolov, J.~R. Ensher, E.~E. Eyler, H. Wang, W.~C. Stwalley, and P.~L.
  Gould, Phys. Rev. Lett. {\bf 84},  246  (2000).

\bibitem{PhysRevLett.94.203001}
J.~M. Sage, S. Sainis, T. Bergeman, and D. DeMille, Phys. Rev. Lett. {\bf 94},
  203001  (2005).

\bibitem{PhysRevLett.101.133005}
F. Lang, K. Winkler, C. Strauss, R. Grimm, and J.~H. Denschlag, Phys. Rev.
  Lett. {\bf 101},  133005  (2008).

\bibitem{deMiranda:2011gd}
M.~H.~G. de~Miranda, A. Chotia, B. Neyenhuis, D. Wang, G. Qu{\'e}m{\'e}ner, S.
  Ospelkaus, J.~L. Bohn, J. Ye, and D.~S. Jin, Nature Physics {\bf 7},  502
  (2011).

\bibitem{PhysRevLett.112.070404}
B. Zhu, B. Gadway, M. Foss-Feig, J. Schachenmayer, M.~L. Wall, K.~R.~A.
  Hazzard, B. Yan, S.~A. Moses, J.~P. Covey, D.~S. Jin, J. Ye, M. Holland, and
  A.~M. Rey, Phys. Rev. Lett. {\bf 112},  070404  (2014).

\bibitem{focus_ultracoldmolecules}
L.~D. Carr and J. Ye, New Journal of Physics {\bf 11},  055009  (2009).

\bibitem{PhysRevLett.114.205302}
J.~W. Park, S.~A. Will, and M.~W. Zwierlein, Phys. Rev. Lett. {\bf 114},
  205302  (2015).

\bibitem{PhysRevLett.116.225306}
S.~A. Will, J.~W. Park, Z.~Z. Yan, H. Loh, and M.~W. Zwierlein, Phys. Rev.
  Lett. {\bf 116},  225306  (2016).

\bibitem{review_molecules2}
B. Gadway and B. Yan, Journal of Physics B: Atomic, Molecular and Optical
  Physics {\bf 49},  152002  (2016).

\bibitem{MolonyPRL}
P.~K. Molony, P.~D. Gregory, Z. Ji, B. Lu, M.~P. K\"oppinger, C.~R. Le~Sueur,
  C.~L. Blackley, J.~M. Hutson, and S.~L. Cornish, Phys. Rev. Lett. {\bf 113},
  255301  (2014).

\bibitem{TakekoshiPRL}
T. Takekoshi, L. Reichs\"ollner, A. Schindewolf, J.~M. Hutson, C.~R. Le~Sueur,
  O. Dulieu, F. Ferlaino, R. Grimm, and H.-C. N\"agerl, Phys. Rev. Lett. {\bf
  113},  205301  (2014).

\bibitem{GoulvenPRL2016}
M. Guo, B. Zhu, B. Lu, X. Ye, F. Wang, R. Vexiau, N. Bouloufa-Maafa, G.
  Qu\'em\'ener, O. Dulieu, and D. Wang, Phys. Rev. Lett. {\bf 116},  205303
  (2016).

\bibitem{Yan:2013fn}
B. Yan, S.~A. Moses, B. Gadway, J.~P. Covey, K.~R.~A. Hazzard, A.~M. Rey, D.~S.
  Jin, and J. Ye, Nature {\bf 501},  521  (2013).

\bibitem{Bloch:2005p988}
I. Bloch, Nature Physics {\bf 1},  23  (2005).

\bibitem{Bloch:2008p943}
I. Bloch, J. Dalibard, and W. Zwerger, Rev. Mod. Phys. {\bf 80},  885  (2008).

\bibitem{white1992}
S.~R. White, Phys. Rev. Lett. {\bf 69},  2863  (1992).

\bibitem{white1993}
S.~R. White, Phys. Rev. B {\bf 48},  10345  (1993).

\bibitem{dmrgbook}
{\em Density Matrix Renormalization - A New Numerical Method in Physics},
  edited by I. Peschel, X. Wang, M. Kaulke, and K. Hallberg (Springer Verlag,
  Berlin, 1999).

\bibitem{daley04}
A.~J. Daley, C. Kollath, U. Schollw\"{o}ck, and G. Vidal, J. Stat. Mech.:
  Theor. Exp.  P04005  (2004).

\bibitem{white04}
S.~R. White and A.~E. Feiguin, Phys. Rev. Lett. {\bf 93},  076401  (2004).

\bibitem{feiguin05}
A.~E. Feiguin and S.~R. White, Phys. Rev. B {\bf 72},  020404(R)  (2005).

\bibitem{manmana:269}
S.~R. Manmana, A. Muramatsu, and R.~M. Noack, AIP Conference Proceedings {\bf
  789},  269  (2005).

\bibitem{noack:93}
R.~M. Noack and S.~R. Manmana, AIP Conference Proceedings {\bf 789},  93
  (2005).

\bibitem{Schollwock:2011p2122}
U. Schollw{\"o}ck, Annals of Physics {\bf 326},  96  (2011).

\bibitem{PhysRevA.84.033619}
A.~V. Gorshkov, S.~R. Manmana, G. Chen, E. Demler, M.~D. Lukin, and A.~M. Rey,
  Phys. Rev. A {\bf 84},  033619  (2011).

\bibitem{tJoriginal1}
P.~W. Anderson, Science {\bf 235},  1196  (1987).

\bibitem{tJoriginal2}
F.~C. Zhang and T.~M. Rice, Phys. Rev. B {\bf 37},  3759  (1988).

\bibitem{auerbach}
A. Auerbach, {\em Interacting Electrons and Quantum Magnetism} (Springer, New
  York, 1994).

\bibitem{dagotto}
E. Dagotto, Rev. Mod. Phys. {\bf 66},  763  (1994).

\bibitem{tJ1977}
K.~A. Chao, J. Spalek, and A.~M. Oles, Journal of Physics C: Solid State
  Physics {\bf 10},  L271  (1977).

\bibitem{Moreno}
A. Moreno, A. Muramatsu, and S.~R. Manmana, Phys. Rev. B {\bf 83},  205113
  (2011).

\bibitem{Schollwock:2005p2117}
U. Schollw{\"o}ck, Rev. Mod. Phys. {\bf 77},  259  (2005).

\bibitem{vidal03}
G. Vidal, Phys. Rev. Lett. {\bf 91},  147902  (2003).

\bibitem{Bedurftig:1998p457}
G. Bed{\"u}rftig, B. Brendel, H. Frahm, and R.~M. Noack, Phys. Rev. B {\bf 58},
   10225  (1998).

\bibitem{White:2002p348}
S.~R. White, I. Affleck, and D.~J. Scalapino, Phys. Rev. B {\bf 65},  165122
  (2002).

\bibitem{BLBQ}
S.~R. Manmana, A.~M. L\"auchli, F.~H.~L. Essler, and F. Mila, Phys. Rev. B {\bf
  83},  184433  (2011).

\bibitem{Barankov2004}
R.~A. Barankov, L.~S. Levitov, and B.~Z. Spivak, Phys. Rev. Lett. {\bf 93},
  160401  (2004).

\bibitem{Natu2013}
S.~S. Natu and E.~J. Mueller, Phys. Rev. A {\bf 87},  053607  (2013).

\end{thebibliography}

\end{document}